\documentclass[11pt]{article}
\usepackage{jcapmod}
\usepackage{tocloft}

\usepackage{framed}
\usepackage{booktabs}
\usepackage[english]{babel}
\usepackage{amsmath,amssymb,amsbsy,amstext, amsthm, simplewick}
\usepackage{hyperref}
\usepackage{graphicx}
\usepackage{amsfonts}
\usepackage{amssymb}
\usepackage{upgreek}
\usepackage{simplewick}
 \usepackage{exscale,relsize}

\usepackage[margin=1cm,labelfont={bf,footnotesize},textfont={footnotesize}]{caption}

\usepackage{colortbl}
\definecolor{lightgreen}{cmyk}{0.2, 0, 0.2, 0.2}
\definecolor{lightgray}{cmyk}{0.1,0.2,0,0.1}
\definecolor{lightgray2}{cmyk}{0.1,0.1,0,0.1}

\makeatletter
\newlength{\apb@width}
\newcommand{\autoparbox}[2][c]{\settowidth{\apb@width}{#2}\parbox[#1]{\apb@width}{#2}}

\makeatother

\numberwithin{equation}{section}

\def\beq{\begin{equation}}
\def\eeq{\end{equation}}

\def\bea{\begin{eqnarray}}
\def\eea{\end{eqnarray}}

\def\d{{\rm d}}
\def\k{\boldsymbol{k}}
\def\x{\boldsymbol{x}}

\def\Mp{M_{\rm pl}}

\def\fnl{f_{\mathsmaller{\rm NL}}}
\def\gnl{g_{\mathsmaller{\rm NL}}}

\RequirePackage{color}

\begin{document}

\begin{titlepage}

\setcounter{page}{1} \baselineskip=15.5pt \thispagestyle{empty}

\bigskip\

\vspace{2cm}
\begin{center}
{\fontsize{20}{28}\selectfont  \sffamily \bfseries Planck-Suppressed Operators}

\end{center}

\vspace{0.2cm}

\begin{center}
{\fontsize{13}{30}\selectfont  Valentin Assassi,$^{\bigstar}$ Daniel Baumann,$^{\bigstar}$ Daniel Green,$^{ \blacklozenge, \clubsuit}$ and Liam McAllister$^{ \spadesuit}$}
\end{center}

\begin{center}

\vskip 8pt
\textsl{$^\bigstar$ D.A.M.T.P., Cambridge University, Cambridge, CB3 0WA, UK}
\vskip 7pt

\textsl{$^ \blacklozenge$
Stanford Institute for Theoretical Physics, Stanford University, Stanford, CA 94305, USA}
\vskip 7pt
\textsl{$^\clubsuit$ Kavli Institute for Particle Astrophysics and Cosmology, Stanford, CA 94305, USA}

\vskip 7pt
\textsl{$^\spadesuit$ Department of Physics, Cornell University, Ithaca, NY 14853, USA}

\end{center}

\vspace{1.2cm}
\hrule \vspace{0.3cm}
{ \noindent {\sffamily \bfseries Abstract} \\[0.1cm]
We show that the recent Planck limits on primordial non-Gaussianity impose strong constraints on light hidden sector fields coupled to the inflaton via operators suppressed by a high mass scale~$\Lambda$.  We study a simple effective field theory in which a hidden sector field is coupled to a shift-symmetric inflaton via arbitrary operators up to dimension five.  Self-interactions in the hidden sector
lead to non-Gaussianity in the curvature perturbations.
To be consistent with the Planck limit on local non-Gaussianity, the coupling to any hidden sector with light fields and natural cubic couplings must be suppressed by a very high scale $\Lambda > 10^5 H$.  
Even if the hidden sector has Gaussian correlations, nonlinearities in the mixing
with the inflaton still lead to non-Gaussian curvature perturbations.  In this case, the non-Gaussianity is of the equilateral or orthogonal type, and the Planck data requires~$\Lambda > 10^2 H$.

\noindent
}
 \hrule

\vspace{0.6cm}

\end{titlepage}

 \tableofcontents

\newpage

\section{Introduction}

The recent results of the Planck satellite \cite{PlanckParameters, PlanckInflation, PlanckNG}
represent a new standard in precision cosmology.
With cosmic variance limited measurements of the temperature anisotropy up to $\ell_{\rm max}  \sim 1500$, the theory of initial conditions is tested at a precision of $\ell_{\rm max}^{-1} \sim 7\times 10^{-4}$.
This has led to a significant improvement in the constraints on the non-Gaussianity of the primordial fluctuations~\cite{PlanckNG}.  These are reported as constraints on the amplitudes of three bispectrum templates---local~\cite{Gangui:1993tt}, equilateral~\cite{Babich:2004gb}, and orthogonal~\cite{Senatore:2009gt}:
\beq
 \fnl^{\rm local} = 2.7 \pm 5.8 \ , \qquad \fnl^{\rm equil} = -42 \pm 75 \qquad {\rm and} \qquad\fnl^{\rm ortho}= -25 \pm 39  \ .   \label{equ:PlanckNG}
\eeq
Recalling that a rough quantitative measure of the amount of non-Gaussianity is $|\fnl| \Delta_\zeta \lesssim 10^{-4}$,
the limits in~(\ref{equ:PlanckNG}) represent a remarkable level of Gaussianity in the primordial perturbations.

It is very well known that limits on non-Gaussianity lead to constraints on the self-interactions of the inflaton.\footnote{In terms of the cutoff scale $\Lambda$ of an effective field theory for the fluctuations of a single inflaton~\cite{Cheung:2007st}, the constraints on $\fnl^{\rm equil}$ and $\fnl^{\rm ortho}$ in (\ref{equ:PlanckNG}) imply that $\Lambda \gtrsim 3 \hskip 1pt H$~\cite{Baumann:2011su}, where $H$ is the inflationary expansion rate. }
However, the requirement that the inflaton generates prolonged accelerated expansion entails further, model-dependent constraints on inflaton self-interactions, which can be stronger than the constraint from (non)observation of the bispectrum.  For example, in slow-roll models
the inflaton~$\Phi$ respects an approximate shift symmetry that constrains self-interactions in the potential to be small and  non-Gaussianity to be unobservable~\cite{Maldacena:2002vr}.
In contrast, self-interactions of
additional fields
$\Sigma$ that couple weakly to the inflaton are much less constrained: $\Sigma$ could have a large cubic coupling\footnote{The conformal symmetry of the inflationary quasi-de Sitter background fixes the three-point function of $\Sigma$ up to an overall normalization~\cite{Creminelli:2011mw}.  We therefore lose no generality by focusing on $\Sigma^3$.}
\beq
{\cal L}_\Sigma \supset - \mu \Sigma^3\ , \label{equ:LS}
\eeq with $\mu\sim H$, and hence have non-Gaussian correlations, without interfering with the slow-roll evolution.

\begin{figure}[h!]
   \centering
       \includegraphics[scale =0.6]{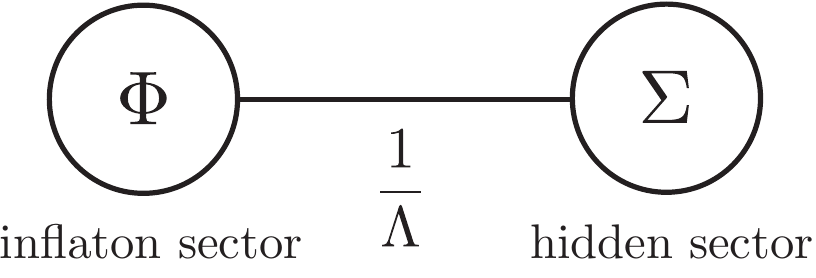}
   \caption{The inflaton sector and the hidden sector mix through irrelevant operators suppressed by the scale $\Lambda$.}
  \label{fig:mixing}
\end{figure}

In this paper, we ask how the Gaussianity of the CMB can be used to place constraints on hidden sector fields.  We consider a hidden sector field $\Sigma$ that is coupled to a shift-symmetric inflaton $\Phi$ via irrelevant operators suppressed by a high scale $\Lambda$.
We construct a general effective field theory~(EFT) involving these fields,
\beq
{\cal L}_{\rm eff}[\Phi,\Sigma] = {\cal L}_\Phi + {\cal L}_\Sigma + {\cal L}_{\rm mix}[\Phi,\Sigma] \ .
\eeq
The leading mixing respecting the shift symmetry is the dimension-five operator
\beq
{\cal L}_{\rm mix}=  - \frac{1}{2} \frac{(\partial \Phi)^2 \Sigma}{\Lambda}\ . \label{equ:LMIX}
\eeq
Such a coupling to the inflaton kinetic term arises rather naturally in ultraviolet (UV) completions of inflation~\cite{BM}.
If $\Sigma$ is light enough\footnote{The presence of hidden sector fields with masses $m \lesssim H$ is not unreasonable from the perspective of UV completions of inflation~\cite{BM}: for example, $\Sigma$ could be a modulus that is massless before supersymmetry breaking.} to be quantum-mechanically active during inflation ($m < \frac{3}{2} H$), self-interactions in the $\Sigma$ sector can be imprinted as non-Gaussianities in the visible curvature perturbations, through the mixing (\ref{equ:LMIX}).
In fact, the theory has two distinct sources of non-Gaussianity: self-interactions in the hidden sector, (\ref{equ:LS}), and nonlinear couplings between the two sectors, (\ref{equ:LMIX}).  As a result, the phenomenology of the model is rather rich, with all three bispectrum shapes probed by Planck realized in different regions of the parameter space.
We will use the Planck bounds on non-Gaussianity to put a lower bound on the scale~$\Lambda$ in (\ref{equ:LMIX}), \beq
\Lambda \, >\, {\cal O}(10^5{\rm -} 10^2)\hskip 1pt H \ ,
\eeq
where the precise numerical coefficient on the r.h.s.~depends on the couplings of the hidden sector.  
A detection of primordial tensors would show that $H \gtrsim 10^{-5} \Mp$, so that $\Lambda$ can exceed the Planck scale.  In this case, the limits on the bispectrum~\cite{PlanckNG} constrain the ultraviolet completion of gravity. 

A similar strategy was first outlined in \cite{Green:2013rd}, for particular classes of strongly-interacting hidden sectors.
In the analysis of \cite{Green:2013rd}, the bispectrum could be computed  only when the mixing between the sectors  was perturbatively small,  so that a region of parameter space was excluded  not by data but by the need for theoretical control.  In the present setting, we will be able to compute the observable signatures for all values of $\Lambda$.

Our approach is analogous to the use of high-precision measurements in particle physics to constrain the scale $\Lambda$ of physics beyond the Standard Model (SM).
There one can place lower bounds on $\Lambda$ that depend sensitively on the effective operators under consideration, ranging from $\Lambda \gtrsim 5\times 10^3$ GeV for precision electroweak measurements~\cite{Skiba:2010xn} to $\Lambda \gtrsim 10^{16}$ GeV for proton decay~\cite{Beringer:1900zz}. 
Moreover, the discovery that neutrinos have finite masses hints at new physics at $\Lambda \sim 10^{15}$ GeV.\footnote{The EFT of the SM contains a unique dimension-five operator involving two leptons and the Higgs, ${\cal L}_{\rm eff}^{\mathsmaller{(5)}} =(LH)(LH)/\Lambda$. For $\Lambda \sim 10^{15}$ GeV, this operator induces neutrino masses of interesting size, $m_\nu \sim 10^{-2}$ eV. }  In this paper, we apply the same philosophy to inflation and use the Planck data to put precision constraints on the scale $\Lambda$ of couplings to hidden sectors.

\vskip 8pt
\noindent
{\it Outline.}---In Section~\ref{sec:EFT}, we construct an effective field theory that couples a shift-symmetric inflaton field to a hidden sector field with strong self-interactions.
We derive the effective action for small fluctuations around a slow-roll background, and we briefly discuss the radiative stability of different parameter regimes.
In Section~\ref{sec:Pheno}, we use both analytical and numerical techniques to study the phenomenology of this theory.
By scanning the complete parameter space of the EFT, we discover that all three bispectrum shapes probed by Planck are realized.
We then show how the Planck limits on local, equilateral and orthogonal non-Gaussianity probe physics at energy scales orders of magnitude above the inflationary expansion rate.
In Section~\ref{sec:SUSY}, we consider the illustrative example of a supersymmetric hidden sector.  We explain how the hidden sector self-interactions, and the corresponding non-Gaussianity, depend on how supersymmetry breaking is communicated to $\Sigma$.  Our conclusions appear in Section~\ref{sec:conclusions}.
Appendix~\ref{sec:Numerical} contains details of the numerical analysis, while Appendix~\ref{app:EFT} provides further discussion of the EFT of the two-scalar system.

Throughout, we will use natural units with $\hbar = c \equiv 1$ and reduced Planck mass $\Mp \equiv 1/\sqrt{8\pi G}.$  Our metric signature is $(-+++)$ and we define $(\partial \Phi)^2 \equiv g^{\mu \nu} \partial_\mu \Phi \partial_\nu \Phi$. 

\section{Effective Theory of Multi-Field Inflation}
\label{sec:EFT}

\subsection{EFT of Background Fields}

Our interest in this paper is the coupling of an approximately shift-symmetric inflaton field $\Phi$ to a hidden-sector field $\Sigma$, which may have significant self-interactions. Mixing operators suppressed by a high scale~$\Lambda$ turn the non-Gaussianity in the hidden sector into observable curvature perturbations.
In this section, we discuss the effective field theory of $\Phi$-$\Sigma$ couplings.\footnote{See~\cite{Baumann:2011nk, Senatore:2010wk, Noumi:2012vr} for an alternative perspective using the effective theory of inflationary fluctuations~\cite{Cheung:2007st}. We note that the derivative expansion in the effective theory of the background fields is different from that of the fluctuations~\cite{Weinberg:2008hq}.}

\vskip 4pt	
We assume that the quantum-corrected inflaton action takes the slow-roll form
\beq
{\cal L}_\Phi = - \frac{1}{2}(\partial \Phi)^2 - V(\Phi) \ .  \label{equ:Lp}
\eeq
The observed (near) scale-invariance of the primordial fluctuations~\cite{PlanckParameters, PlanckInflation} suggests that $\Phi$ respects an approximate shift symmetry, $\Phi \mapsto \Phi + const$.
This constrains self-interactions in the potential to be small and non-Gaussianity to be unobservable~\cite{Maldacena:2002vr}. Large interactions could still come from higher-derivative operators such as $(\partial \Phi)^4/\Lambda^4$~\cite{Creminelli:2003iq}.  However, these operators only produce observable non-Gaussianity if $\Lambda \lesssim (\dot \Phi)^{1/2}$, and in that case the derivative expansion of the EFT cannot be truncated, so that the theory requires a UV completion to be predictive~\cite{Silverstein:2003hf}.\footnote{In this case, the constraints on non-Gaussianity are more usefully organized in terms of an EFT for inflationary fluctuations~\cite{Cheung:2007st}.  The advantage of this approach is that it captures all single-field models, including those with $\Lambda \lesssim (\dot \Phi)^{1/2}$, in the regime of control of a single EFT.}
Although it would be straightforward to generalize our treatment to include higher-derivative kinetic terms, in this paper we will consider only non-Gaussianity from single-derivative couplings.

We take the hidden sector to have the Lagrangian
\beq
{\cal L}_\Sigma = - \frac{1}{2}(\partial \Sigma)^2 - V(\Sigma) \ . \label{equ:Ls}
\eeq
Interactions in the hidden sector are much less constrained than in the inflaton sector, and in particular the self-interactions in the potential $V(\Sigma)$ can be large.  At cubic order in $\Sigma$, the self-interactions of $\Sigma$ are completely characterized by $\mu \Sigma^3$:  higher-derivative operators can be written in terms of $\Sigma^3$ and total derivatives (see footnote 2).
When $\mu \sim H$, the fluctuations of $\Sigma$ have order-one non-Gaussianity.

We characterize the communication between the two sectors in terms of the most general mixing operators allowed in the EFT,
\beq
{\cal L}_{\rm mix}[\Phi,\Sigma] = \sum_I c_I \frac{{\cal O}_I[\Phi,\Sigma]}{\Lambda^{\delta_I - 4}} \ ,
\eeq
where the operators ${\cal O}_I$ are made out of powers of the fields $\Phi$ and $\Sigma$ and their derivatives.
The parameters~$\delta_I$ are the mass dimensions of the operators ${\cal O}_I$, and the $c_I$ are dimensionless Wilson coefficients.
Before imposing that $\Phi$ enjoys an approximate shift symmetry, the leading terms in ${\cal L}_{\rm mix}$ are simply the lowest-dimension operators involving both  $\Phi$ and $\Sigma$.
In Table~\ref{Table:Operators} we list all possible operators constructed from at most first derivatives, up to dimension $\delta=5$.
	
	 \begin{table}[h!]

	\heavyrulewidth=.08em
	\lightrulewidth=.05em
	\cmidrulewidth=.03em
	\belowrulesep=.65ex
	\belowbottomsep=0pt
	\aboverulesep=.4ex
	\abovetopsep=0pt
	\cmidrulesep=\doublerulesep
	\cmidrulekern=.5em
	\defaultaddspace=.5em
	\renewcommand{\arraystretch}{1.6}

	\begin{center}
		\small
		\begin{tabular}{cl}

			\toprule
		Dimension & Operators \\
			\midrule
		\rowcolor[gray]{0.9}{0} &  $V_0$				\\
		\midrule
				\rowcolor[gray]{0.9}{1} &  $\Phi$ , $\Sigma$ . \\
				\midrule
					\rowcolor[gray]{0.9}{2} &  $\Phi^2$ , $\Sigma^2$ , $\Phi \Sigma$ . \\
					\midrule
						\rowcolor[gray]{0.9}{3} &  $\Phi^3$\ , $\Sigma^3$ , $\Phi^2 \Sigma$ , $\Phi \Sigma^2$ . \\
						\midrule
						\rowcolor[gray]{0.9}{4} &  $\Phi^4$\ , $\Sigma^4$ , $\Phi^3 \Sigma$ , $\Phi^2 \Sigma^2$ , $\Phi \Sigma^3$ , \\[-1mm]
						\rowcolor[gray]{0.9}{} &  $\partial_\mu\Phi \partial^\mu \Phi$\ , $\partial_\mu \Sigma \partial^\mu \Sigma$ , $\partial_\mu \Phi \partial^\mu \Sigma$ .  \\
				\midrule
				\rowcolor[gray]{0.9}{5} &  $\Phi^5$\ , $\Sigma^5$ , $\Phi^4 \Sigma$ , $\Phi^3 \Sigma^2$ , $\Phi^2 \Sigma^3$ , $\Phi \Sigma^4$ ,  \\[-1mm]
\rowcolor[gray]{0.9}{} &  $(\partial_\mu \Phi \partial^\mu \Phi) \Phi$ , $(\partial_\mu \Sigma \partial^\mu \Sigma) \Sigma$ ,  $(\partial_\mu \Sigma \partial^\mu \Sigma) \Phi$ ,\\[-1mm]
\rowcolor[gray]{0.9}{}  & $(\partial_\mu \Phi \partial^\mu \Phi) \Sigma$ , $(\partial_\mu \Phi \partial^\mu \Sigma) \Sigma$ , $(\partial_\mu \Phi \partial^\mu \Sigma) \Phi$ . \\[1mm]
 			\bottomrule
		\end{tabular}
	\end{center}
	\vspace{-0.2cm}
	\caption{List of all operators constructed from at most first derivatives, up to dimension 5.
	\label{Table:Operators}}
	\end{table}
	
The requirement of an approximate shift symmetry\footnote{We do not have to commit to whether this symmetry is a fundamental symmetry of the UV theory, an accidental symmetry of the IR description, or simply a consequence of fine-tuning in the quantum corrected effective potential.} $\Phi \mapsto \Phi + const.$ excludes all operators involving $\Phi$ (as opposed to $\partial_\mu \Phi$).
Three mixing operators remain:
\begin{itemize}
\item  $\delta=4$ : \, $\partial_\mu \Phi \partial^\mu \Sigma$

This dimension-four kinetic mixing can be removed by a rigid rotation in field space, $\Phi \mapsto \tilde{\Phi} = \rm{cos}(\theta)\Phi-\rm{sin}(\theta)\Sigma$ and  $\Sigma \mapsto \tilde{\Sigma} =\rm{sin}(\theta)\Phi+\rm{cos}(\theta)\Sigma$.  But as a result $V(\Sigma)=V(\tilde{\Sigma},\tilde{\Phi})$, so that $\tilde{\Phi}$ is not shift-symmetric.
In order for the canonical field  $\tilde{\Phi}$ (with diagonal kinetic term) to be shift symmetric, we will assume that the coefficient of $\partial_\mu \Phi \partial^\mu \Sigma$ is negligibly small.
This is equivalent to imposing the shift symmetry {\it{after}} diagonalizing the kinetic term.
We emphasize that this shift symmetry is motivated by the fact that the primordial fluctuations are observed to be nearly scale-invariant.

\item  $\delta=5$ : \, $(\partial_\mu \Phi \partial^\mu \Sigma) \Sigma$

The dimension-five coupling $(\partial_\mu \Phi \partial^\mu \Sigma) \Sigma$ can be integrated by parts to give\footnote{For equal-time correlation functions, it may not be obvious that a total derivative can be neglected, due to the boundary term.  In the $in$-$in$ formalism, these boundary terms are always associated with equal-time commutators and  (assuming locality) can be removed by a redefinition of the local operators. }
\beq
{\cal L}_{\rm mix}^A \equiv  (\partial_\mu \Phi \partial^\mu \Sigma) \frac{\Sigma}{\Lambda} \ \to\ - \frac{1}{2} \frac{\Box \Phi}{\Lambda} \Sigma^2 \ . 
\label{equ:23}
\eeq
This operator is redundant: using the $\Phi$ equation of motion, $\Box \Phi =  \partial_\Phi V$, we can rewrite it in terms of $ \Phi^m \Sigma^n$.

\item  $\delta=5$ : \, $(\partial_\mu \Phi \partial^\mu \Phi) \Sigma$

A primary goal of this paper is to characterize the effects of the interaction
\begin{equation}
{\cal L}_{\rm mix}^B = - \frac{1}{2}\frac{(\partial \Phi)^2 \Sigma}{\Lambda} \ . \label{equ:Lmix}
\end{equation}
This operator is the dominant source of mixing between fluctuations in the hidden sector and the visible sector.

\end{itemize}

The operator (\ref{equ:Lmix}) contains a tadpole,
\beq
{\cal L}_{\rm mix}^B \, \supset\, \frac{ \dot \Phi^2}{2\Lambda} \hskip 2pt \Sigma\ ,
\eeq
that drives the field $\Sigma$ away from the origin.
We will assume that the potential $V(\Sigma)$ in (\ref{equ:Ls}) stabilizes $\Sigma$ at $\Sigma_0 = const.$ (in Appendix~\ref{app:EFT}, we examine the naturalness of this assumption). Once $\Sigma$ has developed a vev, the interaction~(\ref{equ:Lmix}) induces a correction to the inflaton kinetic term,
\beq
{\cal L}_\Phi = - \frac{1}{2}(1+\kappa) (\partial \Phi)^2 - V(\Phi)\ ,
\eeq
where $\kappa \equiv \Sigma_0/\Lambda$.  We will usually work in the regime where this is a small effect (i.e.~$\kappa \ll 1$).
Even when $\kappa \sim 1$, this correction can be absorbed by a field redefinition, $\tilde \Phi^2 \equiv (1 + \kappa) \Phi^2 $.
In terms of the new field, the mixing term becomes
\beq
{\cal L}_{\rm mix}^B = - \frac{1}{2}\frac{(\partial \tilde \Phi)^2 \Sigma}{\tilde \Lambda} \ ,  \label{equ:Lmix2}
\eeq
where $\tilde \Lambda \equiv (1+\kappa) \Lambda$. From now on we will assume that the rescaling has been performed and drop the tildes.
Moreover, we assume that the potential for $\Phi$ (really $\tilde \Phi$) satisfies the slow-roll conditions, and that the inflaton has a time-dependent background vev $\Phi_0(t)$.

\subsection{EFT of Coupled Fluctuations}

Cosmological observables are sourced by the small fluctuations around the background vevs, i.e.~
\beq
\Phi(t,\x) \equiv \Phi_0(t) + \varphi(t,\x)\qquad {\rm and} \qquad  \Sigma(t,\x) = \Sigma_0 + \sigma(t,\x)\ .
\eeq
We will work in spatially flat gauge, $g_{ij} = a^2 \delta_{ij}$, so that the primordial curvature perturbation is~\cite{Baumann:2009ds}
\beq
\zeta(t,\x) \equiv - \frac{H}{\dot \Phi_0} \varphi(t,\x) \ .
\eeq
At leading order in the slow-roll expansion, the inflaton fluctuations~$\varphi$ are massless\footnote{Expressed in terms of $\zeta$ this statement is exact.} and the mixing between matter and metric fluctuations vanishes.
Eqs.~(\ref{equ:Lp}) and (\ref{equ:Lmix}) then become\footnote{We have dropped a cosmological constant term and two tadpoles proportional to $\dot \varphi$ and $\sigma$. These terms are cancelled by potential terms in eqs.~(\ref{equ:Lp}) and (\ref{equ:Ls}).  This is discussed further in Appendix~\ref{app:EFT}.}
\beq
{\cal L}_\Phi + c_B {\cal L}_{\rm mix}^B = - \frac{1}{2} (\partial \varphi)^2 +\, \rho \hskip 1pt \dot \varphi \sigma - \frac{1}{2 } \frac{(\partial \varphi)^2 \sigma}{\Lambda}\ , \label{equ:29}
\eeq
where we have defined the important mixing parameter
\beq
\rho \equiv \frac{\dot \Phi_0}{\Lambda} \ . \label{equ:rhoDEF}
\eeq
We have absorbed the Wilson coefficient
$c_B$ into the definition of $\Lambda$.
Similarly, eqs.~(\ref{equ:Ls}) and (\ref{equ:23}) give\footnote{The operator ${\cal L}_{\rm mix}^A$ induces
\begin{align}
{\cal L}_{\rm mix}^A = - \frac{1}{2} \frac{\Box \Phi_0}{\Lambda} \sigma^2 - \frac{1}{2}\frac{\Box \varphi }{\Lambda}\left(  \sigma^2 + 2 \Sigma_0 \sigma \right) \ . \nonumber
\end{align}
Using the equations of motion, $\Box \Phi_0 \approx -3 H \dot \Phi_0 = -3 \rho H \Lambda$ and $\Box \varphi =   \rho ( \dot \sigma + 3 H \sigma)$, these terms can be absorbed into the definitions of $m^2$ and $\mu$.}
\beq
{\cal L}_{\Sigma} + c_A {\cal L}_{\rm mix}^A = - \frac{1}{2}(\partial \sigma)^2 - \frac{1}{2} m^2 \sigma^2 -  \mu \sigma^3  + \cdots\ ,
\eeq
where $m^2 \equiv V_0'' - 3 c_A\hskip 1pt (1 - \frac{\Sigma_0}{\Lambda})\rho H$ and $\mu \equiv \frac{1}{3!} V'''_0 +  c_A \frac{\rho H}{\Lambda}$ (dropping contributions that are suppressed by slow-roll parameters).
Up to cubic order, the complete Lagrangian for the coupled $\varphi$-$\sigma$ system is then
\beq
{\cal L}_{\rm eff}[\varphi,\sigma] \ =\ \underbrace{\, - \frac{1}{2} (\partial \varphi)^2 - \frac{1}{2}(\partial \sigma)^2 - \frac{1}{2} m^2 \sigma^2 \, }_{{\cal L}_{0}} \ +\  \rho\hskip 1pt \dot \varphi \sigma \ - \underbrace{\ \frac{1}{2} \frac{(\partial \varphi)^2 \sigma}{\Lambda}  -  \mu \sigma^3 \ }_{{\cal L}_{\rm int}} \ . \ \label{equ:Lstart}
\eeq
In the following, we will study the phenomenology of the Lagrangian~(\ref{equ:Lstart}), both analytically and numerically.

\subsection{Dynamics and Naturalness}
\label{sec:dynamics}

The effective Lagrangian (\ref{equ:Lstart}) contains three independent parameters:  the mass $m$ and the cubic coupling~$\mu$ of the hidden sector field, and the strength $\rho$ of the mixing with the visible sector.
 Depending on the size of the mixing parameter $\rho$ relative to $H$, we  encounter different dynamical regimes.
Throughout our analysis, it will be useful to consider {\it strong mixing} ($\rho \gg H$) and {\it weak mixing} ($\rho \lesssim H$) separately.  These two cases are qualitatively different,  in terms of dynamics, observational signatures and the natural range of parameters.  We can understand the origin of this difference directly from the quadratic Lagrangian,
\beq
{\cal L}_2 = - \frac{1}{2} (\partial \varphi)^2 - \frac{1}{2}(\partial \sigma)^2 - \frac{1}{2} m^2 \sigma^2 + \rho\hskip 1pt \dot \varphi \sigma\ . \label{equ:L2}
\eeq
The equations of motion associated with this Lagrangian are
 \begin{align}
\ddot \varphi + 3 H \dot \varphi + \frac{k^2}{a^2} \varphi &\ =\ - \rho \big[ \dot \sigma + 3 H \sigma \big] \ , \label{EOM1}\\
\ddot \sigma + 3 H \dot \sigma + \left(\frac{k^2}{a^2} + m^2 \right) \sigma  &\ =\ \rho \hskip 1pt \dot \varphi\ , \label{EOM2}
\end{align}
where we have assumed a flat FRW background with scale factor $a(t)$ and Hubble parameter $H = \dot a /a$.  These equations of motion admit WKB-like solutions, where $\varphi \propto \sigma \propto e^{-i \int^t \omega(t') \d t'}$, as long as $\omega \gtrsim H$.  By comparing terms in (\ref{equ:L2}), we see that when $\omega \gg \rho$ the mixing term, $\rho \dot \varphi \sigma$, is negligible compared to $\dot \varphi^2$ and $\dot \sigma^2$ and may therefore be treated as a perturbation.  On the other hand, when $ \omega \ll \rho$, the converse is true and we may neglect the kinetic terms $\dot \varphi^2$ and $\dot \sigma^2$ and describe the dynamics in terms of the mixing term and the gradients.  Which of these terms dominates at horizon crossing, $\omega \sim H$, is a key distinguishing feature of the weak and strong mixing cases, respectively.

\subsubsection{Weak Mixing}

\begin{figure}[h!]
   \centering
       \includegraphics[scale =1.0]{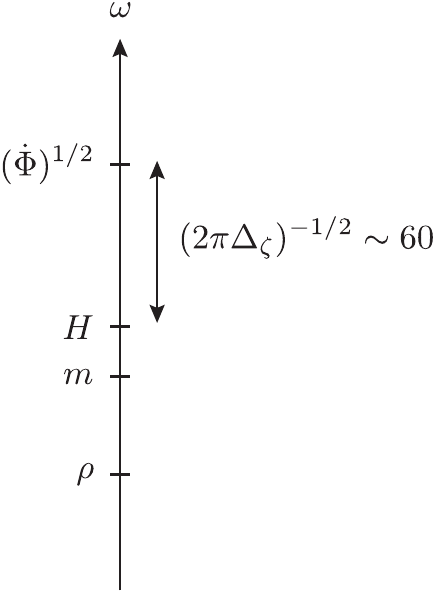}
   \caption{Hierarchies of scales in the weak mixing regime.}
  \label{fig:WeakEnergies}
\end{figure}

In the case of weak mixing, $\rho \lesssim H$, the dominant kinetic terms are the usual ones, $\dot \varphi^2$ and $\dot \sigma^2$.  As such, the dynamics follows a familiar pattern: the relevant scales are the mass $m$ and cubic coupling $\mu$, as well as the mixing scale $\rho$.  The hierarchy of these scales is illustrated in figure \ref{fig:WeakEnergies}.  In order for the fluctuations of $\sigma$ to be generated during inflation (and to be weakly coupled) we will require that $\{m, \mu\} \lesssim H$.
As a result, the full parameter space of this regime is $\{m, \mu, \rho \} \lesssim H$.

\vskip 4pt
Let us briefly discuss whether these conditions are stable under quantum corrections~\cite{Baumann:2011nk}.
First, we estimate the loop corrections to the mass parameter~$m^2$.   The cubic interaction is a soft deformation of the theory and corrects the mass by a finite amount,
\beq
\delta m^2 \sim \mu^2\ .
\eeq  For $\mu < H$, this is an acceptably large correction to the mass.  On the other hand, the loop correction associated with the operator $\frac{1}{\Lambda} (\partial \varphi)^2 \sigma$ is
\beq\label{equ:oneloopmass}
\delta m^2 \sim \frac{\Lambda_{\star}^4}{\Lambda^2} \ ,
\eeq
where $\Lambda_{\star}$ is the UV cutoff of the effective theory of the fluctuations.  We see that the natural value of $m^2$ is sensitive to $\Lambda_{\star}$.
A natural choice for $\Lambda_\star$ is the symmetry breaking scale $(\dot \Phi_0)^{1/2}$~\cite{Cheung:2007st, Baumann:2011su}, i.e.~the scale at which we integrate out the background vevs and focus only on the fluctuations. Letting $\Lambda_\star^2 \sim \dot \Phi_0$, we get
\beq
\delta m^2 \sim \frac{\dot \Phi_0^2}{\Lambda^2} \sim \rho^2 \lesssim H^2\ .
\eeq
Similarly, we can estimate the one-loop correction to the cubic coupling
\beq
\delta \mu \sim \frac{\Lambda_\star^4}{\Lambda^3} \sim \frac{\dot \Phi_0^2}{\Lambda^3} \sim \frac{\rho^2}{\Lambda} \ll H\ .
\eeq
We see that, as long as the cutoff of the EFT of the fluctuations is not far above the symmetry breaking scale, loop corrections do not destabilize the parameters of the EFT.

\subsubsection{Strong Mixing}
\label{ssec:Strong}

In the regime of strong mixing, $\rho \gg H$, the dynamics is controlled by a linear kinetic term $\rho \dot \varphi \sigma$.  When this kinetic term dominates, the scaling behavior of operators is modified and the kinetic terms $\dot \varphi^2$ and $\dot \sigma^2$ become irrelevant.  The quadratic action can be approximated as
\beq
{\cal L}_2 \approx \rho \hskip 1pt  \dot \varphi \sigma - \frac{1}{2} \frac{(\partial_i \varphi)^2}{a^2} - \frac{1}{2} \frac{(\partial_i \sigma)^2}{a^2} - \frac{1}{2} m^2 \sigma^2 \ .  \label{equ:L2Strong}
\eeq
The WKB-like solutions behave as $\omega \sim k^2 /\rho$ at early times.  As a result, if we assign the time coordinate, $t$, a scaling dimension $+1$, then the space coordinate, $x$, will have scaling dimension~$\frac{1}{2}$.
In order to make this scaling manifest, we define $\tilde x^i \equiv  \rho^{1/2} x^i$, $\tilde \varphi \equiv \rho^{-1/4} \varphi$ and $\tilde \sigma \equiv \rho^{-1/4} \sigma$.  In terms of these new variables, the action becomes
\beq\label{equ:strongEFT}
S = \int \d t \hskip 1pt \d^3 \tilde x \ a^3 \left[ \dot{\tilde \varphi} \hskip 1pt \tilde \sigma  - \frac{1}{2} \frac{(\tilde \partial_i \tilde \varphi)^2}{a^2} - \frac{1}{2} \frac{(\tilde \partial_i \tilde \sigma)^2}{a^2} - \frac{1}{2} M \tilde \sigma^2 - \tilde \mu^{1/4} \tilde \sigma^3 -\frac{1}{2} \frac{1}{\tilde \Lambda^{3/4}}\frac{(\tilde \partial_i \tilde \varphi)^2}{a^2} \tilde \sigma\right] \ ,
\eeq
where $M \equiv m^2/\rho$, $\tilde \mu^{1/4} \equiv \mu \rho^{-3/4}$ and $\tilde \Lambda^{3/4} \equiv \Lambda \rho^{-1/4}$.  We have dropped the $\dot \varphi^2 \sigma$ interaction because it is suppressed by additional powers of $\rho$.  In writing this action, we have assumed that the mass term is negligible at the scale $\rho$.  This assumption requires that $M \ll \rho$.  In addition, for the theory to be perturbative at horizon crossing, we also need $\tilde \mu \lesssim H$ and $\tilde \Lambda \gg H$.

\begin{figure}[h!]
   \centering
       \includegraphics[scale =1.0]{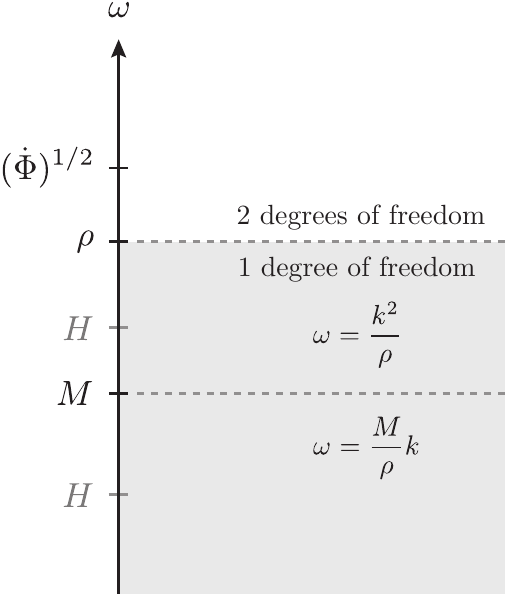}
   \caption{Hierarchies of scales in the strong mixing regime.  The Hubble scale $H$ may be above or below the scale $M=m^2/\rho$.}
  \label{fig:StrongEnergies}
\end{figure}

Despite its appearance, this Lagrangian describes a single degree of freedom. The field $\sigma$ now plays the role of the conjugate momentum of $\varphi$, i.e.~$p_\varphi = \rho \sigma$.  When $\omega \lesssim M$, we can integrate out $\sigma$ (e.g.~by completing the square~\cite{Baumann:2011su}) to produce an effective Lagrangian for $\varphi$ with dispersion relation $\omega = c_s k$, where $c_s = M/\rho$.  For this reason, the phenomenology of this model will depend sensitively on $M/H$, as illustrated in fig.~\ref{fig:StrongEnergies}.

Now that we understand the allowed range of parameters, we should discuss the natural values of these parameters.  The effective theory described by (\ref{equ:strongEFT}) is well-defined up to the energy scale $\rho$ where we are required to include the second degree of freedom.  We assess naturalness of this EFT by computing loops with the UV cutoff $\Lambda_\star \lesssim \rho$.  At one loop, there are three contributions to the mass parameter $M$:
\beq
\delta^{\mathsmaller{(1)}} M \sim \tilde \mu^{1/2} \Lambda_\star^{1/2}  \ , \qquad \delta^{\mathsmaller{(2)}} M \sim \frac{\tilde \mu^{1/4} \Lambda_\star^{3/2}}{\tilde \Lambda^{3/4}} \ ,  \qquad  \delta^{\mathsmaller{(3)}} M \sim \Lambda_\star \left( \frac{\Lambda_\star}{\tilde \Lambda} \right)^{3/2} \ .
\eeq
In writing these expressions, we have neglected additional numerical suppressions such as $(16\pi^2)^{-1}$.  Regardless, we see that $\tilde \mu < H$ and $\tilde \Lambda \gg \rho$ is sufficient to ensure that $M \ll \rho$.  On the other hand, whether $M$ is above or below the Hubble scale depends sensitively on details, including these numerical factors.

The parameters in the effective Lagrangian are also renormalized by loops above the scale $\rho$, including the contributions computed in the previous section.  One might worry that for $\Lambda^2_\star \sim \dot \Phi_0$, eq.~(\ref{equ:oneloopmass}) implies $M \sim \rho$.  These loop corrections could be controlled if the theory is supersymmetric above the scale $\rho$, which would have no impact on the low energy EFT~\cite{Baumann:2011nk}.  Alternatively, including the appropriate factor of $(16 \pi^2)^{-1}$ can be sufficient to create a hierarchy between $\rho$ and $M$.

\section{Non-Gaussian Phenomenology}
\label{sec:Pheno}

The Lagrangian (\ref{equ:Lstart}) contains two distinct sources of non-Gaussianity:
\beq
 {\cal L}_{{\rm int}}^{\mathsmaller{(1)}} = -  \frac{1}{2} \frac{(\partial \varphi)^2 \sigma}{\Lambda}   \qquad {\rm and} \qquad   \  {\cal L}_{\rm int}^{\mathsmaller{(2)}} = -  \mu \sigma^3   \ .
\eeq
These interactions get converted into a bispectrum of primordial curvature perturbations via the quadratic mixing term ${\cal L}_{\rm mix} = \rho\hskip 1pt \dot \varphi \sigma$.
In this section, we will compute the bispectrum as a function of the parameters $m$, $\mu$ and $\rho$ (or $\Lambda$). Our treatment will be numerical, except in the limits of weak mixing~($\rho < H$) and strong mixing ($\rho > H$), where we present analytical results.
The constraints on non-Gaussianity from the Planck satellite can be viewed either as upper limits on $\rho$ or lower limits on $\Lambda$.  The second point of view makes the Planck experiment a probe of high-scale physics.

\subsection{Preliminaries}

For the convenience of the reader, we collect a few basic formulae that are used in the statistical analysis of non-Gaussian perturbations.

The main diagnostic for primordial non-Gaussianity is the bispectrum,
\beq
\langle \zeta_{\k_1} \zeta_{\k_2} \zeta_{\k_3} \rangle = (2\pi)^3\, B_\zeta(k_1,k_2,k_3)\, \delta(\k_1 + \k_2 + \k_3) \ .
\eeq
We will use the {\it in-in} formalism~\cite{Weinberg:2005vy} to compute the bispectrum for the Lagrangian (\ref{equ:Lstart}),
\beq
\langle \zeta_{\k_1} \zeta_{\k_2} \zeta_{\k_3} \rangle(t) = \langle 0 | \left[ \bar T e^{i\int_{t_i}^t \d t' \hat H_{\rm int}(t')}\right]
 \hat \zeta_{\k_1} \hat \zeta_{\k_2} \hat \zeta_{\k_3}(t)  \left[ T e^{-i\int_{t_i}^t \d t' \hat H_{\rm int}(t')}\right] | 0 \rangle\ , \label{equ:inin}
 \eeq
 where $|0\rangle$ is the vacuum of the free theory\footnote{The initial condition will include a small evolution into the imaginary time direction that projects the theory onto the interacting vacuum.  } and $\hat H_{\rm int}[\hat\zeta, \hat \sigma]$ is the interaction Hamiltonian. The
operators $\hat \zeta$ and $\hat \sigma$ are expanded in terms of creation and annihilation operators,
 \beq
 \hat \zeta_{\k}(t) = \zeta_k(t) \hat a_{\k} + \zeta^*_k(t) \hat a^\dagger_{-\k}\ ,
 \eeq
 and similarly for $\hat \sigma_{\k}$.
 The mode functions $\zeta_k(t) \propto \varphi_k(t)$ and $\sigma_k(t)$ are interaction picture fields whose time evolution is determined by the quadratic part of the Hamiltonian, cf.~eqs.~(\ref{EOM1}) and (\ref{EOM2}).
  The bispectrum is computed perturbatively as an expansion in $\hat H_{\rm int}$. Time integrals are performed by analytical continuation into the complex plane (see Appendix~\ref{sec:Numerical}).
 We will compare our results to the three template bispectra used in the Planck analysis~\cite{PlanckNG},
\begin{align}
B_{\rm local} &\equiv \frac{6}{5} \Big( P_1 P_2 + \mbox{\rm 2 perms.} \Big) \ , \label{equ:Bloc} \\
B_{\rm equi} &\equiv \frac{3}{5} \Big( 6\, (P_1^3 P_2^{2} P_3)^{1/3} - 3 P_1 P_2 - 2\, (P_1 P_2 P_3)^{2/3} + \mbox{\rm 5 perms.} \Big) \ , \label{equ:Bequil}\\
B_{\rm ortho} &\equiv  \frac{3}{5} \Big( 18\, (P_1^3 P_2^{2} P_3)^{1/3} - 9 P_1 P_2 - 8\, (P_1 P_2 P_3)^{2/3} + \mbox{\rm 5 perms.} \Big) \ ,\label{equ:Bortho}
\end{align}
where $P_i \equiv P_\zeta(k_i)$ is the power spectrum, $\langle  \zeta_{{\k}_i} \zeta_{{\k}_j}  \rangle =(2\pi)^3 \, P_\zeta(k_i)\,  \delta({\k}_i + {\k}_j)$.  To quantify the degree of correlation between two bispectrum shapes it is convenient to define the following shape function~\cite{Babich:2004gb}:
\beq
S(x_1,x_2) \equiv (x_1 x_2)^2 \, B_\zeta(x_1,x_2,1)\ , \qquad x_i \equiv k_i/k_3\ ,
\eeq
with inner product
\beq
F(S,S') \equiv \int_{\cal V} S(x_1,x_2) S'(x_1, x_2) \, \d x_1 \d x_2\ ,
\eeq
where the integrals are only over physical momenta satisfying $0 \le x_1 \le 1$ and $1-x_1 \le x_2 \le 1$.
Two shapes are highly correlated if their normalized scalar product or `cosine' is close to unity,
\beq
{\cal C}(S,S') \equiv \frac{F(S,S')}{\sqrt{F(S,S) F(S',S')}}\ . \label{equ:cosine}
\eeq
 To facilitate comparison with the constraints in eq.~(\ref{equ:PlanckNG}),
we also compute the amplitude of the bispectrum,
\beq
\fnl \equiv \frac{5}{18} \frac{B_\zeta(k,k,k)}{P_\zeta^2(k)} \ . \label{equ:fnlDEF}
\eeq

\subsection{Spectrum of Non-Gaussianities}
\label{sec:Spectrum}

In Appendix~\ref{sec:Numerical}, we describe in detail our approach to computing the bispectrum numerically.
Both the UV and the IR have to be treated carefully.  In the UV, it is important to make sure that the quantization of the coupled fields is consistent with the equations of motion~(\ref{EOM1}) and (\ref{EOM2}).
In the IR, spurious divergences can appear in the numerical evaluation of the integrals in (\ref{equ:inin}).
The interested reader is referred to the appendix for a detailed description of how we deal with these technical issues.

\begin{figure}[h!]
   \centering
       \includegraphics[scale =0.5]{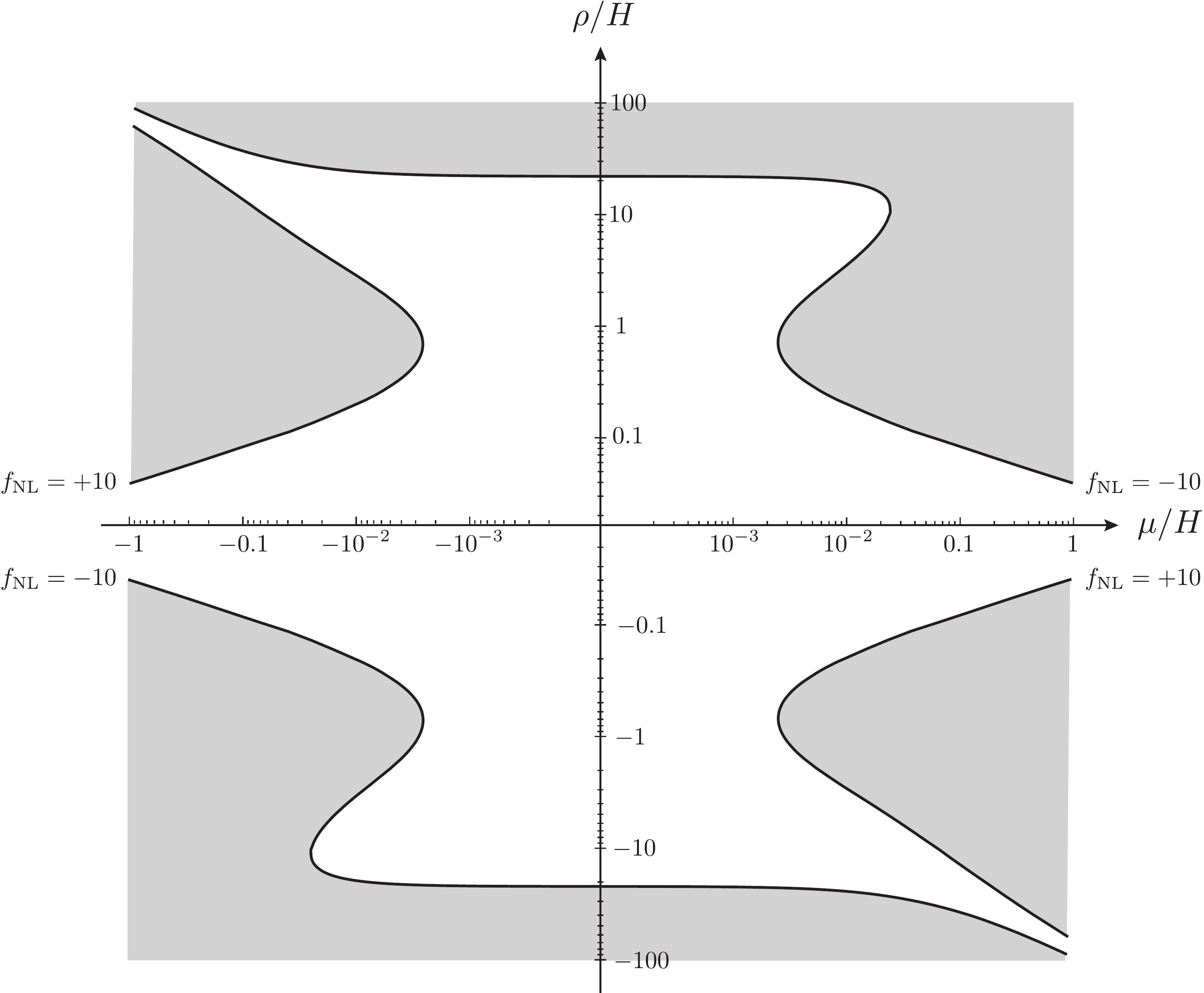}
   \caption{Non-Gaussian phenomenology, for $m=H$. The grey regions correspond to $|\fnl| > 10$.}
  \label{fig:mainRHO}
\end{figure}

\vskip 4pt
Figure~\ref{fig:mainRHO} shows the result of our analysis for the amplitude of the bispectrum, $\fnl$.
For the purpose of illustration, the mass of $\Sigma$
has been fixed to $m=H$. The plot then shows the dependence of $\fnl$ on the parameters $\rho$ and $\mu$. The grey regions denote $|\fnl| > 10$ and are therefore disfavored by the Planck data.\footnote{To be precise about this would require a dedicated likelihood analysis, which is beyond the scope of this paper.}
As we will show, the asymptotic limits of this plot ($\rho \gg H$ and $\rho \ll H$)  can be understood analytically.

\subsubsection{Weak Mixing}
\label{sec:weak}

In the regime of weak mixing, $\rho < H$ (or $\Lambda > \dot \Phi_0/H$), the phenomenology of the $\varphi$-$\sigma$-system was studied in detail in the context of {\it quasi-single-field inflation}~\cite{Chen:2009zp} (see also \cite{Baumann:2011nk}).
We will therefore cite the main results without proof.

\vskip 4pt
\noindent
{\it Mode functions.}---Treating the coupling $\rho \hskip 1pt \dot{\varphi} \sigma$ as a perturbation, the mode functions in the interaction picture are determined by the free field Lagrangian ${\cal L}_0$,
\begin{align}
\varphi'' + 2 {\cal H} \varphi' + k^2 \varphi &= 0 \ , \label{equ:VV}\\
\sigma'' + 2 {\cal H} \sigma' + \big(k^2 + a^2 m^2\big) \sigma &= 0\ . \label{equ:SS}
\end{align}
In a quasi-de Sitter background, $a(\tau) \approx - 1/(H\tau)$, the Bunch-Davies solutions to (\ref{equ:VV}) and (\ref{equ:SS}) are
\begin{align}
\varphi_k(\tau) &= \frac{H}{\sqrt{2 k^3}} (1+ik\tau) e^{-ik \tau} \ , \label{equ:QSFImode}\\
\sigma_k(\tau) &= \sqrt{\frac{\pi}{2}} \frac{H}{\sqrt{2 k^3}}(-k\tau)^{3/2}\, {\rm H}_\nu^{\mathsmaller{(1)}}(-k\tau) \ , \qquad \nu \equiv \sqrt{\frac{9}{4} - \frac{m^2}{H^2}}\ ,
\end{align}
where $\tau$ is conformal time and ${\rm H}_\nu^{\mathsmaller{(1)}}$ is the Hankel function of the first kind.
While $\varphi$ freezes for $|k \tau| \ll 1$ (up to slow-roll corrections), $\sigma$ decays with a rate determined by $m/H$.
The mixing $\rho \hskip 1pt \dot \varphi \sigma$ is treated as part of the interaction Hamiltonian and perturbatively converts the fluctuations $\sigma$ into $\varphi$ (and hence $\zeta$).

\begin{figure}[h!]
   \centering
       \includegraphics[scale =0.35]{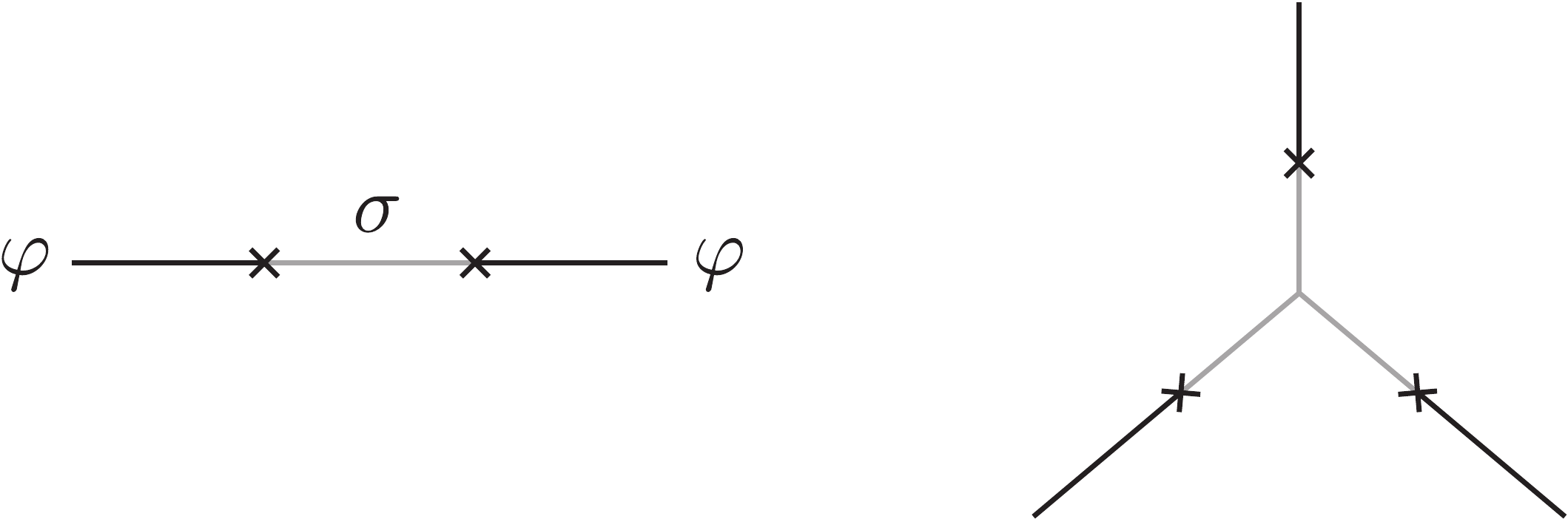}
   \caption{Feynman diagrams describing the conversion of hidden sector fluctuations $\sigma$ into observable fluctuations $\varphi$ (or $\zeta$).}
  \label{fig:QSFI}
\end{figure}

\vskip 4pt
\noindent
{\it Power spectrum.}---The leading-order contribution to the {power spectrum} comes from the inflaton fluctuations, while the mixing with $\sigma$ appears in subleading corrections (see fig.~\ref{fig:QSFI}):
\beq
2\pi \Delta_\zeta = \frac{H^2}{\dot \Phi_0} \left[ 1 + c(m) \frac{\rho^2}{H^2 } \right] \ ,\label{equ:PS}
\eeq
where an explicit expression for the function $c(m)$ can be found in~\cite{Chen:2009zp},
but will not be needed in the following. For the calculation to be under perturbative control
we require $c(m) \rho^2 < H^2$.
Unless $m \ll H$, this is equivalent to the weak mixing condition. For $m \ll H$, the function $c(m) $ becomes large and the condition for perturbative control becomes the stronger one.

\vskip 4pt
\noindent
{\it Bispectrum.}---Since the inflaton fluctuations are Gaussian, the leading contribution to the bispectrum comes from the hidden sector interactions.
Each leg in the interaction ${\cal L}_{\rm int}^{\mathsmaller{(2)}} = -  \mu \sigma^3$ is converted to $\zeta$  via an insertion of $\rho \hskip 1pt \dot \varphi \sigma$.
Using eq.~(\ref{equ:inin}) to evaluate the cubic diagram in~fig.~\ref{fig:QSFI}, one finds
\beq
\fbox{$\displaystyle \fnl = f(m) \cdot \frac{1}{2\pi \Delta_\zeta}  \cdot \frac{\mu}{H} \left( \frac{\rho}{H}\right)^3 $}  \ ,\label{equ:fQSFI}
\eeq
where the function $f(m)$ can be found in~\cite{Chen:2009zp}.  Eq.~(\ref{equ:fQSFI}) explains the $|\rho| < H$ part of fig.~\ref{fig:mainRHO}.

Depending on the mass of the extra field, the shape of the non-Gaussianity interpolates between local ($m < H$) and equilateral ($m \sim H$).  This feature is easy to understand intuitively. When $m \sim H$, the field $\sigma$ decays rapidly on superhorizon scales. The interactions between different modes are therefore suppressed unless they exit the horizon at nearly the same time. This leads to the equilateral shape. On the other hand, if $\sigma$ is nearly massless then modes that exit the horizon at different times in the inflationary history still have non-trivial overlap, and the shape of the bispectrum will be approximately local.

Another important aspect of eq.~(\ref{equ:fQSFI}) is the large prefactor $\Delta_\zeta^{-1}$, which allows for observable levels of non-Gaussianity without violating weak coupling, i.e~for $\mu < H$ and $\rho < H$.
The $\Delta_\zeta^{-1}$ enhancement also admits a straightforward explanation.
For $\rho \sim H$, order-one non-Gaussianity in the $\sigma$-sector (i.e.~$\mu / H \sim 1$) should lead to order-one non-Gaussianity in the $\zeta$-sector. But the relevant measure of the non-Gaussianity of the bispectrum of $\zeta$ is $\fnl \Delta_\zeta$.
 As a result, we should have $\fnl \Delta_\zeta \sim ({\mu/H})({\rho/H})^3$, which is confirmed by the direct computation of (\ref{equ:fQSFI}).

\subsubsection{Strong Mixing}
\label{sec:strong}

When $\mu \ll H$, the dominant source of non-Gaussianity switches from ${\cal L}_{\rm int}^{\mathsmaller{(2)}}$  to  ${\cal L}_{\rm int}^{\mathsmaller{(1)}} = - \frac{1}{2\Lambda} (\partial \varphi)^2 \sigma$.
Observable non-Gaussianity is now generated only if $\Lambda$ is small enough so that $\Lambda < \dot \Phi_0/H$ (and hence $\rho > H$), which is the regime of {strong mixing}.  The dynamics in this region of parameter space is less well understood, although some aspects have been studied in \cite{Baumann:2011su} (see also \cite{Tolley:2009fg, Cremonini:2010ua, Avgoustidis:2012yc, Achucarro:2010da, Shiu:2011qw, Cespedes:2012hu, McAllister:2012am, Burgess:2012dz}). Here, we provide a complete understanding of the phenomenology in the strong mixing limit, as well as numerical results for arbitrary  values of the parameters.

\vskip 4pt
\noindent
{\it Mode functions.}---As we have discussed in \S\ref{sec:dynamics},  for energies below the mixing scale $\rho$, the system reduces to a single effective degree of freedom.
From the Lagrangian (\ref{equ:L2Strong}), we find the equation of motion for $\varphi$,
\beq
 \varphi'' + 2{\cal H} \left[ 1 + \left( 1 + \frac{M}{\Omega_k} \right)^{-1}\right]  \varphi' +  a^2\Omega_k^2  \left(  1 + \frac{M}{\Omega_k} \right) \varphi = 0\ , \label{equ:phiEOM1}
\eeq
where we have defined ${\cal H} \equiv a'/a$, as well as
\beq
\Omega_k(\tau) \equiv \frac{1}{\rho} \frac{k^2}{a^2} \qquad {\rm and} \qquad M \equiv \frac{m^2}{\rho} \ .
\eeq
The solution for $\sigma$ is determined by the  constraint equation
\beq
a\hskip 1pt \sigma =  \frac{ \varphi'}{\Omega_k + M} \ .
\eeq
We see that $\sigma$ is not an independent degree of freedom.
In \S\ref{sec:dynamics}, we also explained that the parameter $M = m^2/\rho$ is the more relevant mass parameter in the strong mixing regime.
For general $M$, eq.~(\ref{equ:phiEOM1}) can only be solved numerically (see Appendix~\ref{sec:Numerical}). However, in the limits $M \gg H$ and $M \ll H$ an analytical understanding is possible:

\begin{itemize}
\item For $M \gg H$, eq.~(\ref{equ:phiEOM1}) becomes
\beq
\varphi'' + 2 {\cal H} \varphi'  + \frac{M}{\rho} k^2 \varphi = 0\ ,  \label{equ:EOM}
\eeq
where we have used that near horizon crossing $\Omega_k \sim H \ll M$. For $\rho > M$, we can identify (\ref{equ:EOM}) as the equation of motion of a scalar with non-trivial {\it sound speed}\,,
\beq
c_s^2 = \frac{M}{\rho} \, .
\eeq
The mode function corresponding to the Bunch-Davies initial state is
\beq
\varphi_k(\tau) = \frac{H}{\sqrt{2 c_sk^3}} \big(1+i c_s k\tau\big) e^{-ic_s k \tau} \ .
\eeq
It is easy to show that this identification with a small-$c_s$ theory extends to the cubic Lagrangian (see~\cite{Baumann:2011su,Tolley:2009fg}).
Both the power spectrum and the bispectrum will therefore be those of a single-field theory with reduced sound speed, $c_s < 1$~\cite{Chen:2006nt}.
\item For $M \ll H$, eq.~(\ref{equ:phiEOM1}) becomes
\beq
\varphi'' + 4 {\cal H}  \varphi' +   \frac{k^4}{\rho^2 a^2} \varphi = 0\ .
\eeq
In de Sitter space, this has the exact solution~\cite{Baumann:2011su}
\beq
 \varphi_k(\tau) = \sqrt{\frac{\pi}{4}} \frac{H}{\rho} \frac{H}{\sqrt{2k^3}}(- k \tau)^{5/2}\ {\rm H}_{5/4}^{\mathsmaller{(1)}}\Big(\frac{1}{2} \frac{H}{\rho}(k \tau)^2 \Big) \ \xrightarrow{k\tau \to 0} \  \frac{2 \Gamma(\frac{5}{4})}{\sqrt{\pi}} \times \frac{H}{k^{3/2}}  \left( \frac{\rho}{H} \right)^{1/4} \ . \label{equ:vp}
\eeq
We see that the inflaton fluctuations are enhanced by a factor of $({\rho/H})^{1/4}$ relative to the canonical slow-roll result.
\end{itemize}

\begin{figure}[h!]
   \centering
       \includegraphics[scale =0.35]{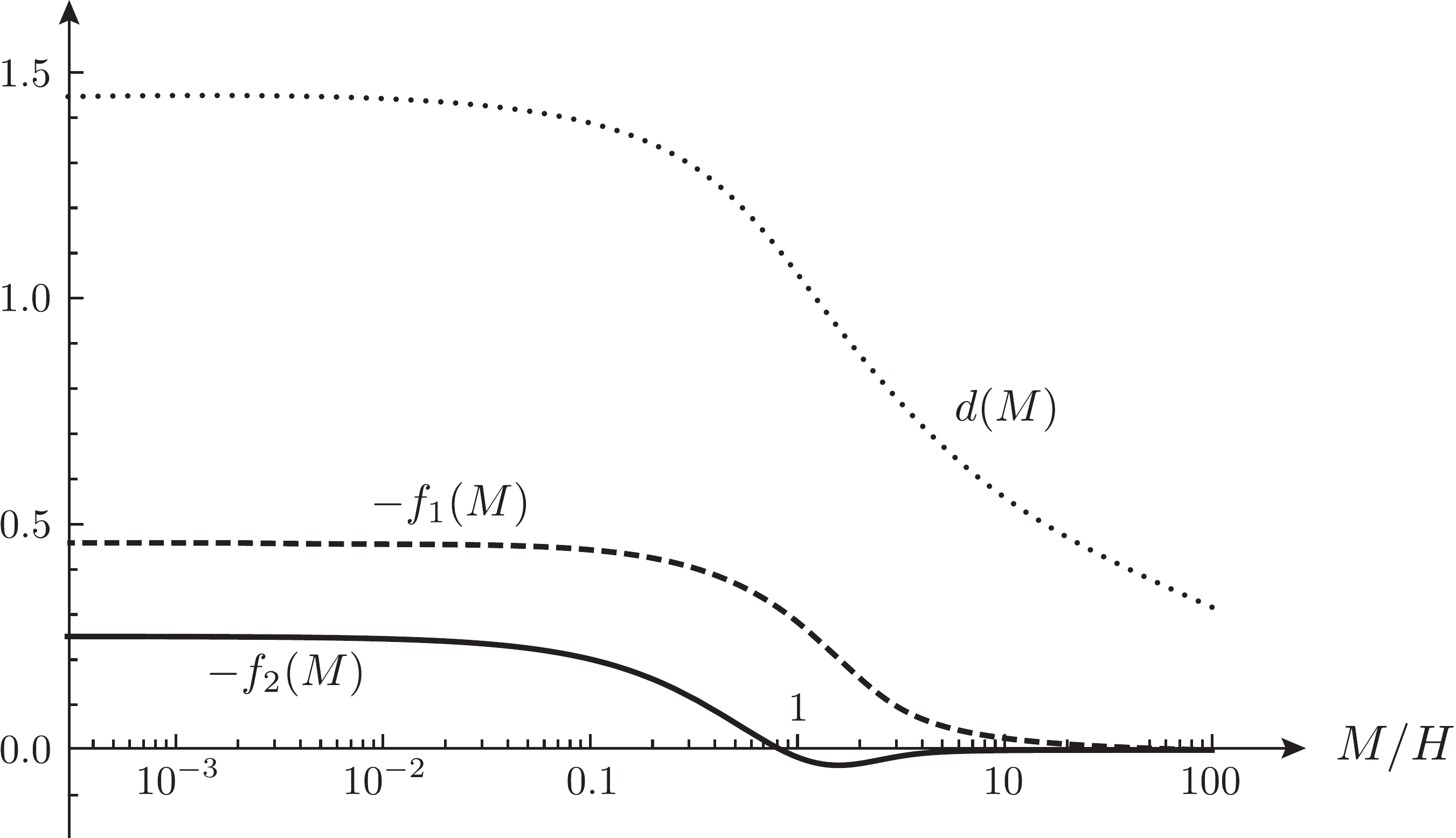}
   \caption{ Numerical computation of $d(M)$ (dotted), $f_1(M)$ (dashed) and $f_2(M)$ (solid). Notice that $f_2(M)$ changes sign near $M=0.8H$.}
  \label{fig:functions}
\end{figure}

\vskip 4pt
\noindent
{\it Power spectrum.}---Given a solution for the mode functions, it is easy to compute the power spectrum of curvature perturbations in the superhorizon limit,
\beq
2 \pi \Delta_\zeta = d(M) \times \frac{H^2}{\dot \Phi_0} \left( \frac{\rho}{H} \right)^{1/4}  \ , \label{equ:DZZ}
\eeq
where the function $d(M)$ is computed numerically and plotted in fig.~\ref{fig:functions}. We have confirmed that our numerical result has the expected asymptotic limit $d(M) \to (H/M)^{1/4}$ for $M \gg H$.\footnote{This follows from $\Delta_\zeta \propto c_s^{-1/2} = (\rho/M)^{1/4}$.}
 
\vskip 4pt
\noindent
{\it Bispectrum from $(\partial \varphi)^2 \sigma$.}---First, we consider the bispectrum generated by the  interaction ${\cal L}_{\rm int}^{\mathsmaller{(1)}} = - \frac{1}{2\Lambda} (\partial \varphi)^2 \sigma$.
The shape is equilateral and the amplitude is
\beq
f_{\mathsmaller{{\rm NL}, (1)}}^{\rm equil} =  f_1(M) \times \frac{\rho}{H}\ . \label{equ:fNL1}
\eeq
The function $f_1(M)$ is computed numerically and plotted in fig.~\ref{fig:functions}. Recall that $M = m^2/\rho$, so for fixed $m$, the function $f_1$ contains a dependence on the mixing parameter $\rho$. We have confirmed that our numerical result has the expected asymptotic limit $f_1(M) \to  - \frac{1}{4} H/M$ for $M \gg H$.\footnote{This follows from $\fnl^{\rm equil} \sim \frac{1}{4} c_s^{-2} = \frac{1}{4}\rho/M$.}
In (\ref{equ:fNL1}), we have factored out $\rho / H$ to match the scaling in the $M \to 0$ limit.  We can understand this scaling from (\ref{equ:strongEFT}) and dimensional analysis,
\beq
f_{\mathsmaller{{\rm NL}, (1)}}^{\rm equil} \sim (2\pi \Delta_\zeta)^{-1} \Big(\frac{H}{\tilde \Lambda}\Big)^{3/4} = \frac{\rho}{H} \ ,
\eeq
where we have defined $\tilde \Lambda^{3/4} = \Lambda \rho^{-1/4}$ as before.

\vskip 4pt
\noindent
{\it Bispectrum from $\sigma^3$.}---A similar analysis for the bispectrum coming from ${\cal L}_{\rm int}^{\mathsmaller{(2)}} = - \mu \sigma^3$ gives
\beq
f_{\mathsmaller{{\rm NL}, (2)}}^{\rm equil} =  \frac{f_2(M)}{2\pi \Delta_\zeta}  \times \frac{\mu}{H} \left( \frac{\rho}{H} \right)^{-3/4} \ ,
\eeq
where $f_2(M)$ is plotted in fig.~\ref{fig:functions}. The numerical result has the correct asymptotic limit $f_2(M) \to (H/M)^{9/4}$ for $M \gg H$.  Again, we have factored out the scaling behavior in the $M \to 0$ limit, which again follows from (\ref{equ:strongEFT}),
\beq
f_{\mathsmaller{{\rm NL}, (2)}}^{\rm equil} \sim (2\pi \Delta_\zeta)^{-1} \Big(\frac{\tilde \mu}{H}\Big)^{1/4} = (2\pi \Delta_\zeta)^{-1} \frac{\mu}{H} \left( \frac{\rho}{H} \right)^{-3/4}  \ ,
\eeq
where $\tilde \mu^{1/4} = \mu \rho^{-3/4}$.

\begin{figure}[h!]
   \centering
       \includegraphics[scale =0.37]{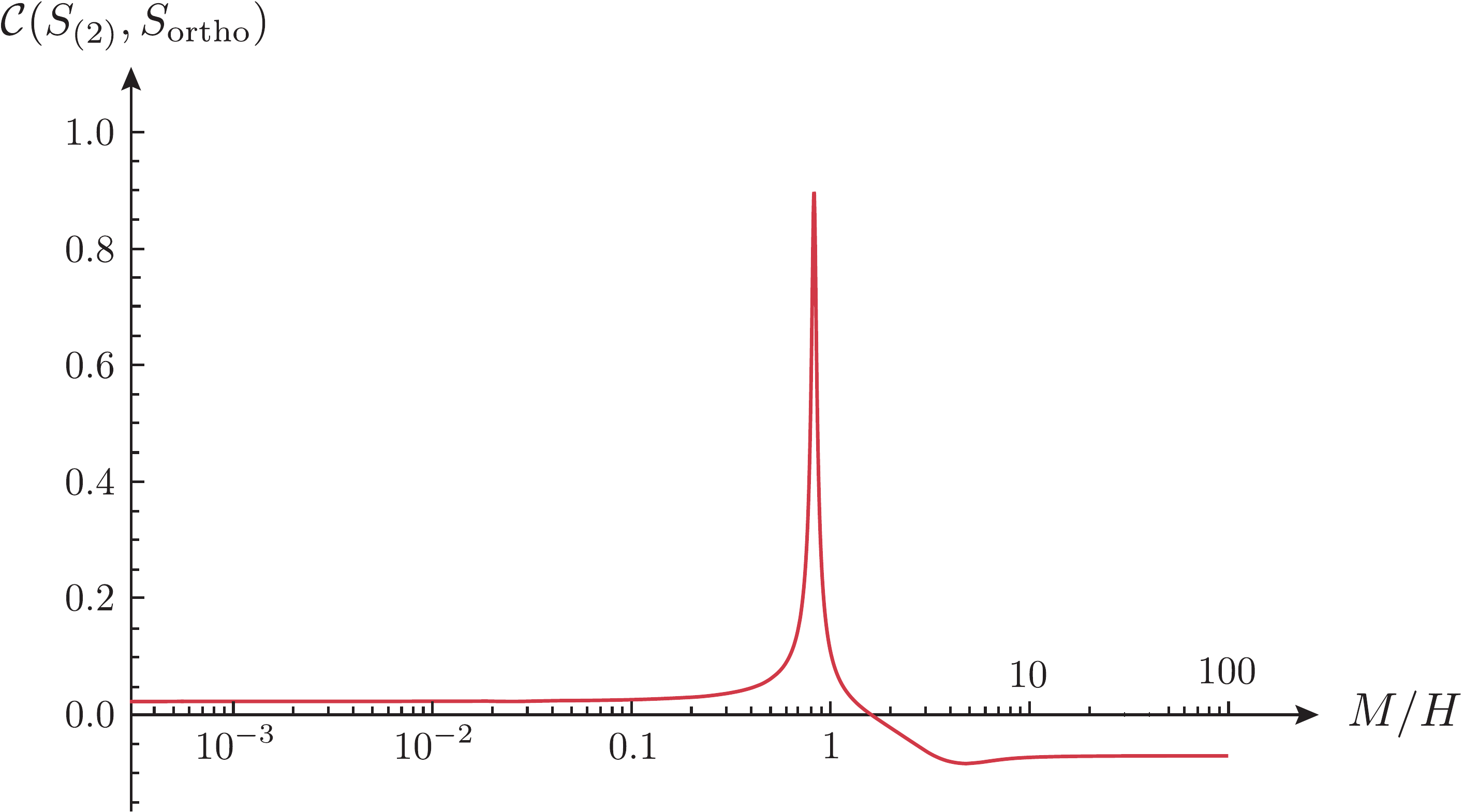}
   \caption{The cosine of the shape $S_{(2)}$ with the orthogonal template shape $S_{\rm ortho}$.}
  \label{fig:Cosine}
\end{figure}

Interestingly, $f_2(M)$---and hence $f_{\mathsmaller{{\rm NL}, (2)}}^{\rm equil}$---changes sign near $M/H = 0.8$.
We interpret this as a cancellation between two large equilateral contributions.
But such a cancellation was precisely what gave rise to the orthogonal shape in~\cite{Senatore:2009gt}, which motivates looking more closely at the shape of the bispectrum near $M/H=0.8$.
We have computed the cosine (\ref{equ:cosine}) between the shape arising from the operator ${\cal L}_{\rm int}^{\mathsmaller{(2)}} = - \mu \sigma^3$ and both the equilateral and the orthogonal templates. The result is shown in fig.~\ref{fig:Cosine}.  We see that the theory indeed realizes both the equilateral and the orthogonal shapes as a function of $M = m^2/\rho$. 
Since the theory also includes the possibility of local non-Gaussianity in the weak mixing limit, our simple model realizes all three standard templates probed by Planck.

\vskip 4pt
\noindent
{\it Complete bispectrum.}---To summarize, the complete bispectrum predicted by the Lagrangian (\ref{equ:Lstart}) in the limit of strong mixing is
\beq
\fbox{$\displaystyle \fnl^{\rm equil} = f_1(M) \cdot \frac{\rho}{H} + f_2(M) \cdot \frac{1}{2\pi \Delta_\zeta} \cdot \frac{\mu}{H}  \left( \frac{\rho}{H} \right)^{-3/4} $} \ . \label{equ:fNLstrong}
\eeq
We note that $\rho$ and $\mu$ can have either sign, so neither the sign of the total $\fnl$, nor the signs of the individual contributions are fixed.  Eq.~(\ref{equ:fNLstrong}) explains the $|\rho| > H$ part of fig.~\ref{fig:mainRHO}.

The result may be written more symmetrically in terms of the parameters in (\ref{equ:strongEFT}),
\beq
(2 \pi \Delta_\zeta) \fnl^{\rm equil} = f_1(M) \times d(M) \cdot \Big(\frac{H}{\tilde \Lambda}\Big)^{3/4} + f_2(M)\Big(\frac{\tilde \mu}{H}\Big)^{1/4} \ . \label{equ:fNLstrong2}
\eeq
This makes it clear that the scaling with energy $H$ is given by $4-\Delta_i$, where $\Delta_i$ is the scaling dimension of the operator that produces the three-point function (see \S\ref{ssec:Strong}).

\begin{figure}[h!]
   \centering
       \includegraphics[scale =0.37]{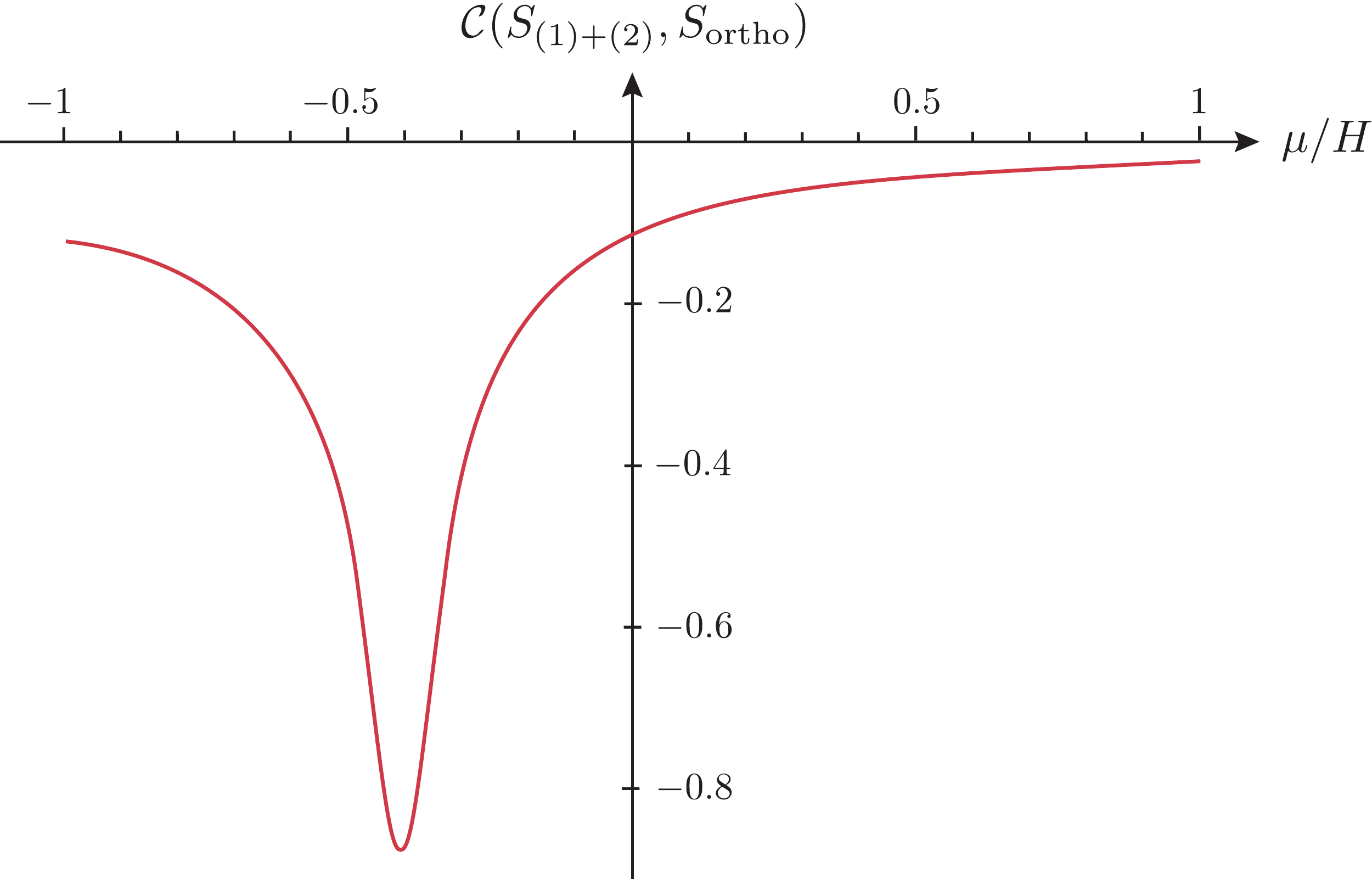}
   \caption{The cosine of $S_{(1)} + S_{(2)}$ with the orthogonal template shape $S_{\rm ortho}$, for $\rho=50H$. }
  \label{fig:Cosine2}
\end{figure}

When $\mu$ and $\rho$ have opposite signs, the two contributions in (\ref{equ:fNLstrong}) can cancel against each other. Naively, this might suggest that we can have large $\mu$ and $\rho$ without producing large non-Gaussianity (see fig.~\ref{fig:mainRHO}). This is a bit misleading: one should recall that the parameter $\fnl$ is defined in the equilateral configuration $k_1=k_2 = k_3$. The two interactions that we are cancelling against each other produce similar, but not identical, shapes.  Away from the equilateral limit, the cancellation will therefore not be perfect. This leads to a non-trivial  bispectrum that is constrained by observations, although the $\fnl$ parameter suggests otherwise.
Again, we can diagnose this by computing the cosine with the orthogonal template. The result is shown in fig.~\ref{fig:Cosine2}: we find an orthogonal component near the point in parameter space where the two equilateral shapes cancel.

\subsection{Probing High-Scale Physics}
\label{sec:hidden}

\begin{figure}[h!]
   \centering
       \includegraphics[scale =0.45]{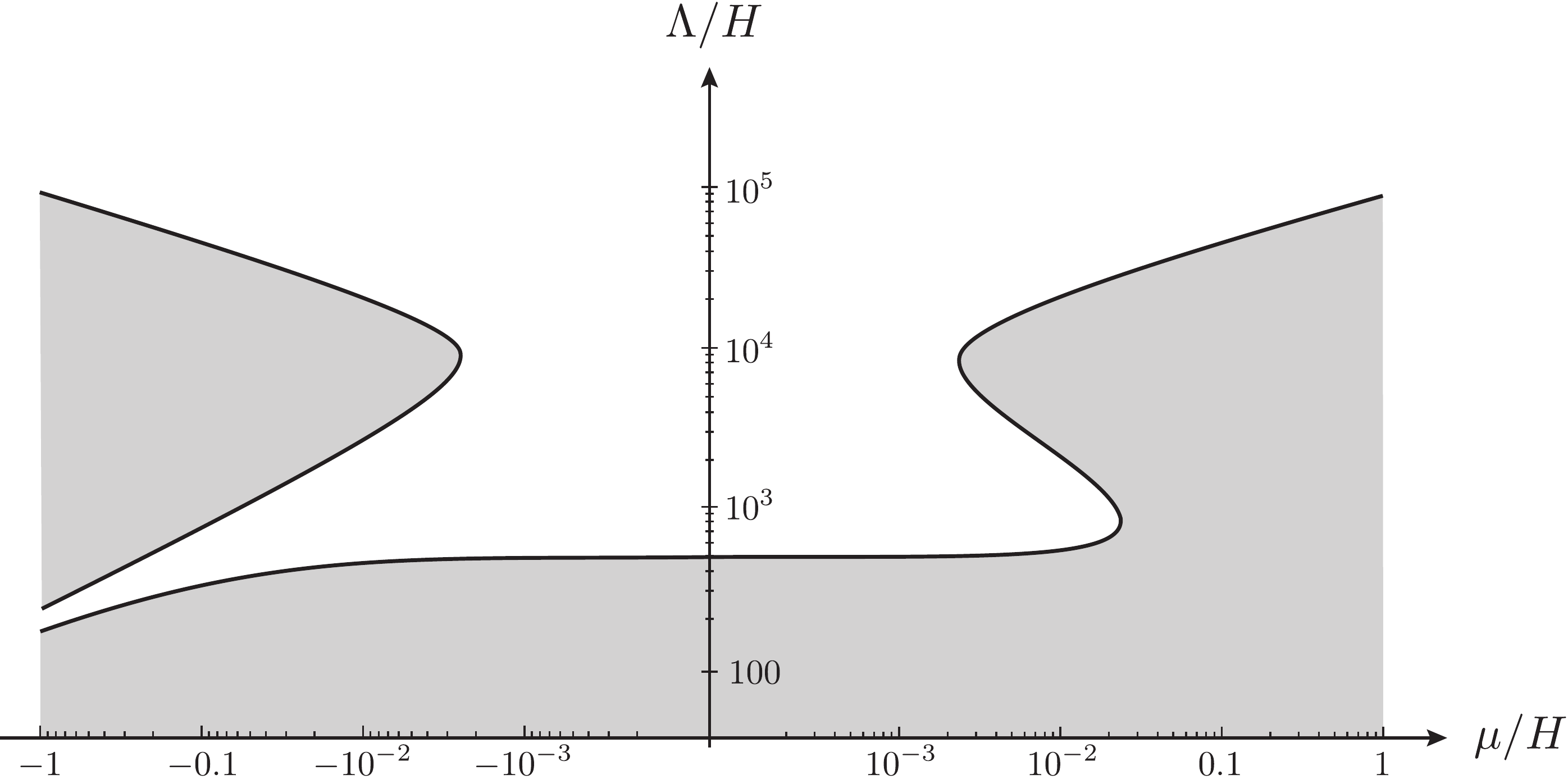}
   \caption{Constraints from Planck on the scale of new physics.}
  \label{fig:mainLAMBDA}
\end{figure}

Fig.~\ref{fig:mainLAMBDA} shows the non-Gaussian predictions of the effective theory expressed in terms of
the suppression scale $\Lambda$ of the dimension-five operator~(\ref{equ:Lmix}), cf.~(\ref{equ:rhoDEF}).
Again we show the region $|\fnl| > 10$ for the case $m=H$.
This indicates how the Planck limits~(\ref{equ:PlanckNG}) constrain high-scale physics.

\vskip 4pt
\noindent
{\it Non-Gaussian hidden sector.}---We get the strongest constraint if the hidden sector is order-one non-Gaussian ($|\mu| \sim H$).
In the weak mixing limit, we use (\ref{equ:PS}) to replace $\dot \Phi_0$ in (\ref{equ:rhoDEF}) and write the bispectrum amplitude~(\ref{equ:fQSFI}) as
\beq
\fnl = \frac{f(m)}{2\pi} \frac{1}{\Delta_\zeta^4} \frac{\mu}{H} \left(\frac{H}{2\pi \Lambda} \right)^3  \ .
\eeq
Solving for $\Lambda$, we obtain
\beq
 \Lambda \, =\, 9 \times 10^4 \left( \frac{|\fnl|}{10}\right)^{-1/3} \left( \frac{|\mu|}{H} \right)^{1/3}\, H \ , \label{equ:L1}
\eeq
where the precise numerical coefficient depends on $m$.
For $m < H$, the scale $\Lambda$ is bounded by Planck's limit on local non-Gaussianity~\cite{PlanckNG}, $|\fnl^{\rm local}| < 10$, while for $m \sim H$ we should impose the limit on equilateral non-Gaussianity, $|\fnl^{\rm equil}| < 75$. A conservative bound on the scale of new physics is therefore
\beq
\Lambda \, \gtrsim\, {\cal O}(5) \times 10^4 \, \left( \frac{|\mu|}{H} \right)^{1/3} \, H\ .
 \label{equ:L3}
\eeq
For $\mu \sim H$, this constrains physics at scales many orders of magnitude above the Hubble scale.

\vskip 4pt
\noindent
{\it Planck-suppressed operators.}---To express (\ref{equ:L3}) as a limit on $\Lambda$ in terms of an absolute energy scale, we need additional observational input.
Specifically, a detection of primordial tensor fluctuations would relate the inflationary expansion rate $H$ to the Planck scale $\Mp$.
In terms of the tensor-to-scalar ratio $r$, we can write
\beq
H = \pi\, \Delta_\zeta \, \sqrt{\frac{r}{2}} \, M_{\rm pl} \sim 10^{-5} \left( \frac{r}{0.01}\right)^{1/2}\, \Mp\ ,
\eeq
and (\ref{equ:L3}) becomes
\beq
\Lambda \, \gtrsim\, 0.5 \, \left( \frac{|\mu|}{H} \right)^{1/3}\, \left( \frac{r}{0.01}\right)^{1/2}\,  \Mp \ .
\eeq
Hence, seeing tensors ($r \gtrsim 0.01$) 
would put $\Lambda$ close to the Planck scale, for couplings to hidden sectors that have order one non-Gaussianity ($|\mu| \sim H$).

\vskip 4pt
\noindent
{\it Gaussian hidden sector.}---If the hidden sector is Gaussian ($|\mu| \ll H$), the constraint on high-scale physics comes from the non-linearity of the mixing operator in the regime of strong mixing. We use (\ref{equ:DZZ}) to replace $\dot \Phi_0$ in (\ref{equ:rhoDEF}) and write (\ref{equ:fNL1}) as
\beq
\fnl =  \frac{\tilde f_1(M) }{\Delta_\zeta^{4/3}} \left( \frac{H}{2\pi \Lambda} \right)^{4/3}  \ ,
\eeq
where $\tilde f_1(M) \equiv f_1(M) \, d^{\hskip 1pt 4/3}(M)$.
The limit $M \ll H$ is particularly interesting, since the signal then only depends on the ratio of $\Lambda$ and $H$.
Solving for $\Lambda$, we obtain
\beq
\Lambda \, =\, 5 \times 10^2 \left( \frac{|\fnl| }{10}\right)^{-3/4} \, H\ .
\eeq
Since Planck constrains equilateral non-Gaussianity at the level $|\fnl^{\rm equil}| < 75$, we conclude that $\Lambda > 110 \hskip 1pt H$.
This limit does not depend on  assumptions about the strength of the self-interactions in the hidden sector---in fact, $\Sigma$ could have purely Gaussian correlations---but it does assume that the hidden sector field is light enough to contribute to the curvature perturbations:  namely,  we must have $M \ll \rho$.\footnote{Recalling that $M = m^2/\rho$ and  that  the mixing is strong when $\rho \gg H$,  we see  that $\Sigma$  can be light enough to contribute through strong mixing ($m^2 \ll \rho^2$)  even if $m \gg H$.  In other words,  fluctuations of $\Sigma$  decouple from the curvature perturbations for $m > \rho$, rather than for $m > \frac{3}{2}H$  as in the weak  mixing case.}
When there are nontrivial self-interactions in the hidden sector, the bound becomes stronger: see fig.~\ref{fig:mainLAMBDA}.

\section{Supersymmetric Hidden Sectors}
\label{sec:SUSY}

We obtained the strongest constraints on high-scale physics, $\Lambda > 10^5 H$, if the  hidden sector field $\Sigma$ was light enough  to be quantum-mechanically active during inflation ($m \lesssim \frac{3}{2}H$) and had a cubic coupling obeying $\mu \gtrsim H$. As we now explain, these conditions are very naturally met  if supersymmetry is spontaneously broken during inflation
\cite{Baumann:2011nk}.
On the other hand, we will  also point out that if the hidden sector containing $\Sigma$ is {\it{sequestered}} from the source of supersymmetry breaking, $\Sigma$  can naturally have nearly Gaussian correlation functions.
In this case, only our weaker---but universal---constraint, $\Lambda > 10^2 H$, applies.

\subsection{Local Non-Gaussianity from Generic Soft SUSY Breaking}

A supersymmetrized version of the inflationary theories discussed in this paper can be modeled  in terms of three  sets of  superfields: a spurion $X$, an  inflaton superfield $\Phi$,  and a hidden sector  superfield $\Sigma$.   The spurion $X$  is taken to be the sole source of supersymmetry breaking.  We assume that this SUSY breaking is communicated to $\Phi$ and $\Sigma$ only through gravity, i.e.~via Planck-suppressed interactions. On the other hand, $\Phi$ and $\Sigma$ can have direct couplings suppressed by a lower scale, such as the mixing term (\ref{equ:Lmix}).  The spontaneous breaking of SUSY by $X$ induces soft terms in both the inflaton sector and the hidden sector.
This situation is analogous to the generation of soft terms in the supersymmetric Standard Model through the spontaneous breaking of supergravity.\footnote{In the application to the MSSM, $\Sigma$ would stand for the `visible' sector, while $X$ is called the `hidden' sector. Here, we have two hidden sectors, one that breaks SUSY ($X$) and one that receives non-Gaussian soft terms ($\Sigma$).}  A complete characterization of SUSY-breaking soft terms in effective supergravity can be found in~\cite{Kaplunovsky:1993rd}.
Here, we only review the basic conclusions relevant for our analysis.

\begin{figure}[h!]
   \centering
       \includegraphics[scale =0.6]{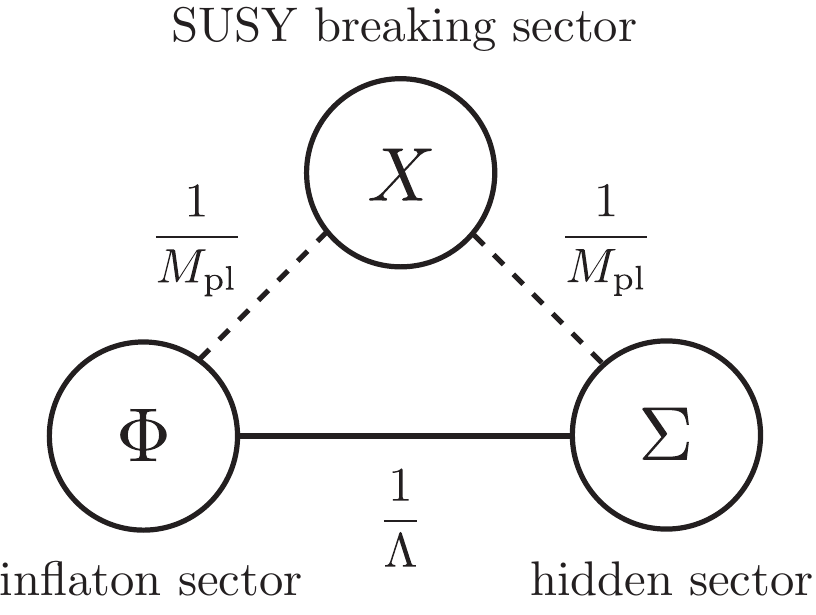}
   \caption{Couplings between the superfields in the supersymmetric effective theory of inflation.}
  \label{fig:SUSY}
\end{figure}

\vskip 4pt
\noindent
{\it Generic soft terms.}---We assume that SUSY is exact at high energies, but becomes {spontaneously} broken at low energies. In the low-energy EFT, the SUSY breaking is characterized by soft terms and the details of the breaking mechanism decouple. In the model of fig.~\ref{fig:SUSY}, these soft terms can be computed in terms of ${\cal N}=1$ supergravity input~\cite{Kaplunovsky:1993rd}. For generality, we now allow both $X$ and $\Sigma$ to be collections of superfields, i.e.\ $X^i$ and $\Sigma^I$, with $i=1, \dots, n_X$ and $I=1,\dots, n_\Sigma$.
We take the superpotential $W$ to be a general holomorphic function of $X$ and $\Sigma$, and expand around $\Sigma=0$:
\beq
W(X,\Sigma) = \hat W(X) + \frac{1}{2} \tilde \mu_{IJ}(X) \Sigma^I \Sigma^J + \frac{1}{3} \tilde Y_{IJL}(X) \Sigma^I \Sigma^J \Sigma^L + \cdots \ .
\eeq
Famously, $W$ is not renormalized at any order in perturbation theory \cite{Grisaru}, so that symmetry structures imposed on the tree-level superpotential receive at most nonperturbatively small corrections.
The K\"ahler potential $K$ takes the following form
\beq
K(X, \bar X, \Sigma, \bar \Sigma) = \hat K(X,\bar X) + Z_{I \bar J}(X,\bar X) \Sigma^I \bar \Sigma^{\bar J} + \left(\frac{1}{2} H_{IJ}(X,\bar X) \Sigma^I \Sigma^J + h.c. \right) + \cdots \ .
\eeq
All couplings in $K$ receive perturbative corrections.
The scalar potential for the spurion $X$ is
\beq
V(X,\bar X) = \hat K_{i\bar \jmath} F^i \bar F^{\bar \jmath} - 3 e^{\hat K} |\hat W|^2\ , \label{vxx}
\eeq
where $\hat K_{i\bar \jmath} \equiv \partial_i \partial_{\bar \jmath} \hat K$ and $\bar F^{\bar \jmath} \equiv e^{\hat K/2} \hat K^{\bar \jmath i} (\partial_i \hat W + \hat W \partial_i \hat K)$. We assume that the potential has a stable minimum with at least one non-zero F-term $\langle F^i \rangle \ne 0$,  so that SUSY is spontaneously broken. The order parameter that measures the SUSY breaking is the gravitino mass,
\beq
m_{3/2} \equiv e^{\langle \hat K \rangle/2} \big|\hat W(\langle X \rangle)\big| \ . \label{m32}
\eeq
Integrating out the spurion, one arrives at an effective theory for the field $\Sigma$,
\beq
W_{\rm eff}(\Sigma) = \frac{1}{2}\mu_{IJ} \Sigma^I \Sigma^J + \frac{1}{3} Y_{IJL} \Sigma^I \Sigma^J \Sigma^L \ ,
\eeq
where $\mu_{IJ} \equiv e^{\langle \hat K\rangle/2} \langle \tilde \mu_{IJ}\rangle + m_{3/2} \langle H_{IJ} \rangle - \langle \bar F^{\bar \imath} \partial_{\bar \imath} H_{IJ} \rangle$ and $Y_{IJL} \equiv e^{\langle\hat K\rangle/2} \langle \tilde Y_{IJL} \rangle$. This induces the following soft SUSY breaking terms for $\Sigma$,\footnote{For simplicity, we only show the result for $Z_{I\bar J} = \delta_{I \bar J}$. The complete answer can be found in~\cite{Kaplunovsky:1993rd}. }
\beq
V_{soft}(\Sigma) = m_{I \bar J}^2 \Sigma^I \bar \Sigma^{\bar J} + \left( \frac{1}{3} A_{IJL} \Sigma^I \Sigma^J \Sigma^L + \frac{1}{2} B_{IJ} \Sigma^I \Sigma^J + h.c. \right) \ ,
\eeq
where
\begin{align}
m_{I \bar J}^2 &\equiv m_{3/2}^2 \delta_{I \bar J} \ , \quad A_{IJL} \equiv F^i D_i Y_{IJL} \ , \quad {\rm and} \quad
B_{IJ} \equiv F^i D_i \mu_{IJ} - m_{3/2}\hskip 1pt \mu_{IJ} \ ,
\end{align}
with $D_i Y_{IJL} \equiv \partial_i Y_{IJL} + \frac{1}{2} (\partial_i \hat K) Y_{IJL}$ and $D_i \mu_{IJ} \equiv \partial_i \mu_{IJ} + \frac{1}{2} (\partial_i\hat K) \mu_{IJ}$.
We see that the generic scale of all soft terms is set by the gravitino mass:
\beq
m_{I\bar J}^2 \sim m_{3/2}^2\ , \quad A_{IJL} \sim m_{3/2} Y_{IJL}\ , \quad {\rm and} \quad B_{IJ} \sim m_{3/2} \mu_{IJ}\ .
\eeq

The final step  is to relate $m_{3/2}$  to the scale of inflation,  in order to  compare the soft terms to the Hubble scale~$H$.
The precise relation is model-dependent, but  reasonably general estimates are possible.
We assume that  instabilities of the moduli---for example, the decompactification instability of a string compactification---are controlled by barriers in the F-term potential.  Comparing (\ref{vxx})  and (\ref{m32}), we see that the size of the barriers is set by $m_{3/2}^2$  (in units with $\Mp=1$).
With the further assumption  that the  curvature of the moduli potential around  its minimum  is dictated by the scales appearing in $W$ and $K$  (rather than being much larger  than these scales  as a result of fine-tuned cancellations, cf.\ \cite{Kallosh:2004yh}), then the  F-term energy  can be at most of order the barrier height, $ \hat K_{i\bar \jmath} F^i \bar F^{\bar \jmath} \lesssim {\rm few}\times m_{3/2}^2$:  otherwise  the moduli become destabilized.  Since the F-term energy also sets the scale of inflation,  we conclude that  in generic configurations,\footnote{This relationship was first emphasized in \cite{Kallosh:2004yh}.   Constructing natural models with $m_{3/2} \ll H$  remains a model-building challenge despite some dedicated efforts, cf.\ e.g.\ \cite{He:2010uk,Conlon:2008cj},  which  suggests that $m_{3/2} \sim H$ could be a relatively robust  property of inflationary models in supergravity.} the scale of soft terms in SUSY inflation is the Hubble scale $H$.
Thus, the masses and the cubic couplings of the hidden sector fields are quite generally at least of order~$H$.
Of course,  mass terms and cubic couplings are allowed by supersymmetry, and  the supersymmetric  masses could be large enough  to prevent $\Sigma$ from fluctuating during inflation.   When the supersymmetric  contributions to the masses and couplings are small---for example, if $\Sigma$  is a modulus,  with a vanishing potential in the supersymmetric limit---then the soft terms control the dynamics.
 
Comparing to our  analysis of the non-Gaussianity  in terms of the parameters $m$ and $\mu$,  we conclude that if the self-interactions of $\Sigma$  are produced by generic soft supersymmetry breaking, then $m \sim \mu \sim H$,  and the dominant source of non-Gaussianity is  self-interactions of $\Sigma$.  The constraint on the scale of the dimension-five mixing operator is then $\Lambda > 10^5 H$.

\subsection{Equilateral Non-Gaussianity from Sequestered SUSY Breaking}

The  results of the previous section hold for  generic soft symmetry breaking, for which all soft terms are of order the gravitino mass $m_{3/2}$.  If instead the soft terms in the  $\Sigma$ sector are small compared to $m_{3/2}$, we say that $\Sigma$ is {\it{sequestered}} from the source of supersymmetry breaking.   Assuming further that $\Sigma$  does not have large supersymmetric interactions---as above, $\Sigma$  could be a modulus---we have $\mu \ll H$, and $\Sigma$  has nearly Gaussian correlation  functions.   As we have shown, such Gaussian hidden sectors imprint equilateral or orthogonal non-Gaussianity in the curvature perturbations as a result of non-linearities in the mixing with $\Phi$.
In short, we conclude that the dominant  channel for a sector of sequestered moduli to generate non-Gaussianity is by mixing with the inflaton.

It is therefore  worth asking whether sequestering of a hidden sector is  plausible: is it guaranteed by reasonable symmetry assumptions, or in the context of an ultraviolet theory?  Let us address these points in turn, in supergravity and in string theory.

\vskip 4pt
\noindent
{\it Sequestering in supergravity and in string theory.}---Randall and Sundrum~\cite{Randall} originally observed that sequestering, in the form of vanishing tree-level soft terms,  follows from the
separability conditions\footnote{Weaker conditions suffice to get partial suppression of the soft terms \cite{Blumenhagen:2009gk,BergMarsh}.}
\begin{align}
f(X,\Sigma) &= f(X,\bar{X}) + f(\Sigma,\bar{\Sigma})  \label{seq1}  \ ,\\
W(X,\Sigma) &= W(X) + W(\Sigma)\ ,   \label{seq2}
\end{align} where $f \equiv e^{-K/3}$.
To understand whether these conditions are plausible, it is useful to  recognize that sequestering
amounts to the suppression of gravity-mediated interactions between the visible sector and the SUSY-breaking  sector, {\it{beyond}} the degree of suppression  determined by the Planck  scale.
In particular, the K\"ahler  potential coupling
\begin{equation}
\frac{c}{\Mp^2} \int d^4\theta\, X^{\dagger}X \Sigma^{\dagger} \Sigma \label{xxss}
\end{equation}  must have $c \ll 1$.
Because the coupling (\ref{xxss}) is  suppressed by the Planck scale,  a proper justification for $c\ll 1$
requires an ultraviolet completion of gravity.
Unsurprisingly, the condition $c \ll 1$ is rather difficult to justify  through symmetry arguments  in four-dimensional effective quantum field theory.\footnote{The only complete four-dimensional  argument \cite{Luty2},  known as {\it{conformal sequestering}},  assumes that the hidden sector is strongly coupled and  nearly superconformal  over a large range of energies,  and as such is dual to an extra dimension.}  For this reason we turn to extra-dimensional arguments  for sequestering, in supergravity and in string theory.

Before proceeding, we remark that the  condition (\ref{seq2}) might appear easy to justify,  because if the tree-level superpotential is separable, then perturbative nonrenormalization \cite{Grisaru} extends superpotential separability to all orders in perturbation theory.  However, separability at the nonperturbative level does not generally follow, and so superpotential separability is a reasonable assumption only when nonperturbative superpotential terms can be neglected \cite{BergMarsh}.  In fact, there are explicit examples\footnote{Concretely, a visible sector residing on D-branes can receive  large soft terms  through  threshold corrections to a gaugino condensate superpotential  generated on a  distant stack of D7-branes \cite{BergMarsh,BergConlon}.} in string theory in which nonperturbative separability is violated \cite{BergConlon}.

The original  proposal for sequestering of supersymmetry breaking \cite{Randall} relied on locality in an extra dimension: the conditions (\ref{seq1}) and (\ref{seq2}) were  shown to hold  when the supersymmetry-breaking sector and the `visible'\footnote{In the literature on sequestering,  this is the visible sector containing the  Standard Model,  but for the present discussion it is actually the hidden sector containing $\Sigma$.} sector communicated only via  higher-dimensional gravity, with no other modes propagating in the bulk.
Because a full  ultraviolet completion is  desired, one should  determine whether  the assumption of a barren  extra dimension is justifiable in a compactification of string theory,  i.e.~does spatial separation  imply sequestering in  string compactifications?  In general compactifications, spatial isolation alone  does {\it{not}} suffice for sequestering \cite{Anisimov},  because light fields  propagating across the compactification  mediate  interactions that  induce large soft terms (cf.~\cite{Kachru2}).
However, sequestering can be achieved if the visible sector and the hidden sector are separated along a warped extra dimension,  e.g.\ if  these sectors reside on D-branes at opposite ends of a warped throat region \cite{Kachru}:  this is the gravity dual of conformal sequestering \cite{Luty2}.

To summarize the above,  the plausibility of sequestering should be determined in an ultraviolet-complete theory. In string theory, sequestered sectors can arise on  D-branes in warped regions,  though one must  carefully check that nonperturbative superpotential couplings do not spoil sequestering.

\vskip 4pt
\noindent
{\it Properties of a sequestered hidden sector.}---A natural question now is whether the particular structures  that ensure sequestering,  particularly warping, can have any consequences beyond the smallness of the scalar soft terms.   That is, do the sequestered scenarios arising in string constructions  suggest different signatures from those described above in the $\Phi$-$\Sigma$ model?
A general observation is that if warping  is responsible for sequestering,  then the dual approximately conformal field theory  will  be strongly coupled,  with large anomalous dimensions.   Non-Gaussianity from  such a conformal hidden sector  was characterized in \cite{Green:2013rd}, from the perspective that a conformal hidden sector is a reasonable possibility a priori.  Here, we are pointing out in addition that in the context of spontaneously broken supersymmetry,  one has two broad choices---generic soft breaking, leading to non-Gaussianity from self-interactions in the hidden sector,  and sequestered breaking,  with non-Gaussianity from mixing with the inflaton---and the latter  scenario plausibly corresponds to a nearly conformal hidden sector.
This provides further motivation for considering the  theories described in \cite{Green:2013rd}.

\section{Conclusions}
\label{sec:conclusions}

We have shown that the Planck limits on non-Gaussianity~\cite{PlanckNG} imply constraints on hidden sector fields~$\Sigma$ coupled to the inflaton $\Phi$ by nonrenormalizable operators suppressed by a scale $\Lambda$.  We constructed a general EFT for $\Phi$ and $\Sigma$, assuming that $\Phi$ was invariant under a shift symmetry, but imposing no such restriction on $\Sigma$.  The leading mixing between the two sectors comes from the dimension-five operator
\beq
- \frac{1}{2}\frac{(\partial \Phi)^2 \Sigma}{\Lambda} \ .  \label{kineticcoup}
\eeq
Non-Gaussianity in the curvature perturbations arises from two distinct sources: self-interactions in the hidden sector and nonlinear couplings between the two sectors.  When the correlations of $\Sigma$ have order-one non-Gaussianity---which is the case for natural-size cubic couplings---then hidden sector self-interactions dominate, and the curvature perturbations acquire non-Gaussianity of the local or equilateral type, depending on the mass of $\Sigma$.  If instead $\Sigma$ has a small cubic coupling and hence Gaussian correlations, the dominant non-Gaussianity in the curvature perturbations arises from nonlinearities in the $\Sigma$-$\Phi$ coupling, and is of the equilateral or orthogonal type.  Our simple two-field EFT therefore realizes all three bispectrum shapes probed by the Planck satellite.

Although  our analysis did not rely on assumptions about ultraviolet physics,  it is worth pointing out that the effective theory  described here is very natural from the perspective of string theory.   The key ingredients  are an inflaton field; one or more additional scalar fields  that are light enough to fluctuate during inflation---for example,  moduli that acquire masses $m \sim H$ after supersymmetry breaking---and field-dependent kinetic couplings  of the form (\ref{kineticcoup}).   All three  ingredients are commonplace in flux compactifications~\cite{Douglas:2006es, BM}.

Our  results  demonstrate that the Planck limits on non-Gaussianity constrain high-scale physics.  Operators that mix a hidden sector involving light scalars with the inflaton must be suppressed by at least $\Lambda > 10^2 H$.  When the cubic couplings in the hidden sector are of order $H$, the limit is significantly stronger, $\Lambda > 10^5 H$.  Furthermore, a detection of primordial tensors~\cite{Baumann:2008aq}, $r > 0.01$, would place severe constraints on non-Gaussian hidden sectors, by pushing the lower bound on $\Lambda$  almost to the Planck scale:
\beq
\Lambda \, \gtrsim\, 0.5 \, \left( \frac{|\mu|}{H} \right)^{1/3}\, \left( \frac{r}{0.01}\right)^{1/2}\,  \Mp \ .
\eeq
The bispectrum results of Planck would then imply constraints on Planck-suppressed couplings to hidden sectors.
Such  constraints  would provide a powerful selection principle for ultraviolet completions of inflation,  because in known constructions---such as string compactifications  in the  supergravity regime---the effective cutoff scales,  set by the Kaluza-Klein and string scales, are well below the Planck scale.   Indeed,  the  presence of a cutoff  that is parametrically below $\Mp$  is a hallmark of  theoretically controllable ultraviolet completions  of gravity.   We find it  noteworthy that a detection of primordial tensors  would sharply constrain  the existence of light  hidden sector scalars in any such ultraviolet completion.

\subsubsection*{Acknowledgements}

We thank Peter Adshead, Xingang Chen, Xi Dong, Anatoly Dymarsky, Eiichiro Komatsu, Andrew Liddle, Hans-Peter Nilles, Enrico Pajer, Hiranya Peiris, Kris Sigurdson, Eva Silverstein, Yi Wang, Christof Wetterich,
and Matias Zaldarriaga for helpful discussions. D.B.~thanks Eva Silverstein for suggesting the title.
D.B.~and V.A.~gratefully acknowledge support from a Starting Grant of the European Research Council (ERC STG grant 279617).  The research of D.G.~is supported in part by the Stanford ITP and by the U.S. Department of Energy contract to SLAC no.\ DE-AC02-76SF00515.
The research of L.M.~is supported by the NSF under grant PHY-0757868.
D.B.~thanks the Bethe Center for Theoretical Physics, Bonn and the KITP, Santa Barbara for hospitality.

\newpage
\appendix
\section{Details of the Numerical Analysis}
\label{sec:Numerical}

In this appendix, we describe in detail the numerical analysis that produced the results of Section~\ref{sec:Pheno}.

\vskip 4pt
The starting point is the Hamiltonian of the coupled $\varphi$-$\sigma$ system
\beq
H \ =\ \underbrace{ \frac{1}{2} \dot \varphi^2 + \frac{1}{2a^2} (\partial_i \varphi)^2 + \frac{1}{2}\dot \sigma^2 + \frac{1}{2a^2} (\partial_i \sigma)^2 + \frac{1}{2} m^2 \sigma^2}_{{H}_0} \ -\ \rho\hskip 1pt \dot \varphi \sigma \ + \ \underbrace{\frac{1}{2} \frac{(\partial \varphi)^2 \sigma}{\Lambda}   +   \mu \sigma^3}_{{H}_{\rm int}} \ . \label{equ:Hstart}
\eeq
The quadratic part, $H_0 - \rho\hskip 1pt \dot \varphi \sigma$, implies the equations of motion
\begin{align}
\varphi'' + 2 {\cal H}  \varphi' + k^2\varphi &\ =\ - \rho a \big[  \sigma' + 3 {\cal H} \sigma \big] \ , \label{equ:A2}\\
 \sigma'' + 2{\cal H} \sigma' + \left(k^2 + m^2 a^2 \right) \sigma  &\ =\ \rho a \hskip 1pt \varphi'\ , \label{equ:A3}
\end{align}
where ${\cal H} \equiv a'/a$ and primes are derivatives with respect to conformal time $\tau$.
The solutions to these equations determine the mode functions in the interaction picture.  The effects of the two cubic interactions,
\beq
 {H}_{\rm int}^{\mathsmaller{(1)}} =  \frac{1}{2} \frac{(\partial \varphi)^2 \sigma}{\Lambda}  \qquad {\rm and} \qquad  {H}_{\rm int}^{\mathsmaller{(2)}} =   \mu \sigma^3    \ , \label{equ:Hint}
\eeq
are treated perturbatively.
 Throughout this appendix, we approximate the spacetime as de Sitter with scale factor
\beq
a(\tau) = - \frac{1}{\tau}\ ,
\eeq
in units where $H \equiv 1$.

\subsection{Weak Mixing Approximation}

We refer the reader to \cite{Chen:2009zp} for a detailed discussion of the weak mixing regime ($\rho \ll H$).
In this limit the analysis simplifies because we can treat the mixing as a perturbative interaction. The interaction picture mode functions satisfy uncoupled equations of motion whose solutions are known analytically, at least in the quasi-de Sitter approximation.
The bispectrum associated with the cubic self-interaction $\sigma^3$ gets converted to a bispectrum of curvature perturbations $\zeta$ via three insertions of the quadratic mixing interaction (cf.~fig.~\ref{fig:QSFI}).
In the $in$-$in$ formalism, both the cubic interaction vertex and the insertions are integrated over time.
The integrals have to be performed numerically. Results can be found in~\cite{Chen:2009zp,Assassi:2012zq}.
Our treatment of the strong and intermediate mixing regimes~($\rho \gtrsim H$) is new and will be described in the remainder of this appendix.

\subsection{Strong Mixing Approximation}

Parts of the strong mixing regime were studied analytically in~\cite{Baumann:2011su}.  Here, we develop a numerical treatment that covers the entire range of parameters.

\subsubsection{Mode Functions}

In \S\ref{ssec:Strong} and \S\ref{sec:strong}, we described the dynamics of the model in the strong mixing regime.
We showed that the system reduces to a single degree of freedom whose evolution we can determine numerically.

\vskip 4pt
\noindent
{\it Equation of motion.}---For $\rho \gg H$, we can drop $\varphi'' +2 {\cal H}\varphi'$ on the l.h.s.~of (\ref{equ:A2}) and $ \sigma'' + 2{\cal H}  \sigma'$ on the l.h.s.~of (\ref{equ:A3}).  The mode function $\varphi_k(\tau)$ then satisfies
\beq
\varphi_k'' - \frac{2}{\tau}\left(1+\frac{(k\tau)^2}{(k\tau)^2+m^2}\right)\varphi'_k+\frac{k^2}{\rho^2}\Big((k\tau)^2+m^2\Big)\varphi_k =0\ .
\eeq
It is convenient to rescale the time coordinate, $u \equiv k\tau/\sqrt{\rho}$, and write
\beq
\varphi_k(\tau) =\frac{\rho^{1/4}}{k^{3/2}}\cdot\tilde\varphi\left(u \right)\ , \label{equ:rescale} 
\eeq
where $\tilde\varphi(u)$ satisfies
\beq
\tilde \varphi'' - \frac{2}{u}\left(1+\frac{u^2}{u^2+M}\right)\tilde\varphi'+\left(u^2+M\right)\tilde\varphi=0\ , \qquad M\equiv\frac{m^2}{\rho}\ . \label{equ:varEOM}
\eeq
We should think of $\tilde\varphi$ as the solution corresponding to $k=\rho=1$. The solutions for general $k$ and $\rho$ then simply follow from eq.~(\ref{equ:rescale}). The solutions to eq.~(\ref{equ:varEOM}) have to be found numerically, except in the special case $M=0$, where the answer can be written in terms of Hankel functions~\cite{Baumann:2011su},
\beq
\tilde \varphi(u) = A (-u)^{5/2} \, {\rm H}_{5/4}^{\mathsmaller{(1)}}(\tfrac{1}{2}u^2) + B (-u)^{5/2}\,  {\rm H}_{5/4}^{\mathsmaller{(2)}}(\tfrac{1}{2}u^2)\ , \qquad M=0\ .
\eeq

\vskip 4pt
\noindent
{\it Initial conditions.}---In order to define the initial conditions, it is convenient to write
$q = a^2 \varphi$.
 (This removes the friction term.)
Moreover, at early times, $k\tau \to - \infty$, we can expand the equation of motion in powers of the small ratio $m^2/(k \tau)^2$. At lowest non-trivial order, we find
\beq
q_k'' + \underbrace{\frac{k^2}{\rho^2} (k\tau)^2 \left( 1 + \frac{m^2}{(k \tau)^2} \right)}_{\omega_k^2(\tau)} q_k = 0\ .
\eeq
This equation has the
WKB solution 
\beq
\lim_{k \tau \to - \infty} q_k(\tau) = \frac{A_k}{\sqrt{\omega_k(\tau)}} \, e^{-i \int \omega_k(\tau) \, \d \tau}\ ,  \label{equ:WKB}
\eeq
where $\omega_k(\tau) \approx - \frac{k^2}{\rho} \tau - \frac{1}{2} \frac{M}{\tau}$.
The constant $A_k \equiv \frac{1}{\sqrt{2}} \frac{k}{\rho}$ is chosen so that the mode functions satisfy the Wronskian normalization $q_k q_k^*{}' - q_k^* q_k' = i \frac{k^2}{\rho^2}$~\cite{Baumann:2011su}.
Eq.~(\ref{equ:WKB}) then implies
\beq
\lim_{u \to - \infty} \tilde \varphi(u) = \frac{1}{\sqrt{2}} \left(- u\right)^{3/2} \, e^{\frac{i}{2} u^2} e^{\frac{i}{2} M \ln (- u)} \ . \label{equ:EARLY}
\eeq

\vskip 4pt
\noindent
{\it Wick-rotated solutions.}---In order to improve the convergence of the bispectrum integral in the UV, we consider Wick-rotated mode functions.  The WKB solution~(\ref{equ:EARLY}) is suppressed at early times if we choose the Wick rotation $u \to e^{i \frac{\pi}{4}} u$.
We define Wick-rotated mode functions as
\beq
\tilde \varphi_w(u) \equiv \tilde \varphi(e^{i \frac{\pi}{4}} u)\ .
\eeq
Since the mode functions are not known analytically, the Wick rotation has to be implemented at the level of the equation of motion. In particular, along the Wick-rotated integration contour,
eq.~(\ref{equ:varEOM}) becomes
\beq
\tilde\varphi_w''(u) - \frac{2}{u}\left(1+ \frac{u^2}{u^2-iM}\right)\tilde\varphi_w'(u)- \left( u^2 - iM\right)\tilde\varphi_w(u)=0\ ,
\eeq
with initial condition
\beq
\lim_{u \to - \infty} \tilde \varphi_{w}(u) = \frac{1}{\sqrt{2}} \left(- u \right)^{3/2} \, e^{- \frac{1}{2} (u^2-iM\ln(-u))} e^{-  \frac{\pi}{8} M}\ .
\eeq

\subsubsection{Power Spectrum}

The superhorizon limit of the power spectrum of curvature perturbations is
\begin{align}
(2\pi)^2 \Delta_\zeta^2 = \frac{2 k^3}{\dot \Phi^2_0} |\varphi_k(0)|^2 \equiv d^{\hskip1pt 2}(M) \cdot   \frac{H^4}{\dot \Phi^2_0} \left( \frac{\rho}{H}\right)^{1/2} \ ,
\end{align}
where  in the final equality we have restored explicit factors of $H$ and defined
\beq
d^{\hskip1pt 2}(M) \equiv  2|\tilde\varphi(0)|^2\ .
\eeq

\subsubsection{Bispectra}

The cubic interactions in eq.~(\ref{equ:Hint}) induce a bispectrum for the inflaton fluctuations,
\beq
\big\langle\varphi_{\k_1}\varphi_{\k_2}\varphi_{\k_3}\big\rangle \equiv (2\pi)^3\, B_\varphi(k_1,k_2,k_3)\, \delta(\k_1+\k_2+\k_3)\ .
\eeq
In what follows, we will use the notation
\beq
\varphi_{w,k}(v) \equiv \frac{1}{k^{3/2}}\cdot\tilde\varphi_{w}(kv)\ .
\eeq
\vskip 4pt
\noindent
{\it Bispectrum from $(\partial \varphi)^2\sigma$.}---The interaction ${H}_{\rm int}^{\mathsmaller{(1)}} =  \frac{1}{2} \frac{(\partial \varphi)^2 \sigma}{\Lambda}$ gives 
\beq
B_\varphi^{\mathsmaller{(1)}}(k_1,k_2,k_3)  = \frac{\rho}{\Lambda}\, {\cal I}^{\mathsmaller{(1)}}(k_1,k_2,k_3)\ ,
\eeq
with
\begin{align}
{\cal I}^{\mathsmaller{(1)}}(k_1,k_2,k_3) &\ =\ - {\rm Re}\Bigg[ie^{i\tfrac{\pi}{4}}\int_{-\infty}^0\frac{\d v}{v}\,\frac{k_3^2-k_2^2-k_1^2}{M-i(k_3v)^2} \nonumber \\
&\hskip 30pt [\varphi_{w,k_1}(0)\varphi_{w,k_1}^*(v)] [\varphi_{w,k_2}(0)\varphi_{w,k_2}^*(v)] [\varphi_{w,k_3}(0)\varphi_{w,k_3}'^*(v)]+2\ {\rm perms.}\Bigg]\ .
\end{align}
Here, we have ignored the contribution from $(\dot \varphi)^2 \sigma$, which is suppressed by powers of $H/\rho$.

\vskip 4pt
\noindent
{\it Bispectrum from $\sigma^3$.}---The interaction ${H}_{\rm int}^{\mathsmaller{(2)}} =   \mu \sigma^3$ gives 
\beq
B_\varphi^{\mathsmaller{(2)}}(k_1,k_2,k_3)   =\mu\,{\cal I}^{\mathsmaller{(2)}}(k_1,k_2,k_3)\ ,
\eeq
with
\begin{align}
{\cal I}^{\mathsmaller{(2)}}(k_1,k_2,k_3)  &\ =\  12\, {\rm Re}\, \Bigg[ie^{3i\frac{\pi}{4}}\int_{-\infty}^0\frac{\d v}{v}\,\frac{\varphi_{w,k_1}(0)\varphi_{w,k_1}'^*(v)}{M-i(k_1v)^2}\frac{\varphi_{w,k_2}(0)\varphi_{w,k_2}'^*(v)}{M-i(k_2v)^2}\frac{\varphi_{w,k_3}(0)\varphi_{w,k_3}'^*(v)}{M-i(k_3v)^2}\Bigg]\ .
\end{align}

\vskip 4pt
\noindent
{\it Full bispectrum.}---The full bispectrum for the primordial curvature perturbations is
\begin{align}
B_\zeta(k_1,k_2,k_3) &\ =\ -\left( \frac{H}{\dot \Phi_0}\right)^3 \bigg[ B_\varphi^{\mathsmaller{(1)}} + B_\varphi^{\mathsmaller{(2)}} \bigg]  \ .
\end{align}
Using eq.~(\ref{equ:fnlDEF}), we compute the amplitude of the bispectrum
\beq
\fnl= f_1(M) \cdot \frac{\rho}{H} +f_2(M) \cdot \frac{1}{2\pi \Delta_\zeta} \frac{\mu}{H}  \left( \frac{\rho}{H} \right)^{-3/4} \ ,
\eeq
where
\beq
f_1(M) \equiv -\frac{5}{18}\,\frac{{\cal I}^{\mathsmaller{(1)}}(1,1,1)}{|\tilde\varphi(0)|^4}\qquad {\rm and} \qquad
f_2(M) \equiv  -\sqrt{2} \times \frac{5}{18}\,\frac{{\cal I}^{\mathsmaller{(2)}}(1,1,1)}{|\tilde\varphi(0)|^3}\ .
\eeq
Plots of the functions $f_1(M)$ and $f_2(M)$ are shown in fig.~\ref{fig:functions}.
 
\subsection{General Treatment}

Finally, we present a general analysis that is valid for all values of $\rho$, including the intermediate mixing regime $\rho \sim H$. Unlike the weak mixing regime, the interaction picture mode functions are not decoupled. Unlike the strong mixing regime, we cannot ignore half of the solutions.
In this case, a numerical analysis seems unavoidable.  The methodology that we will employ is a generalization of the treatment of~\cite{vanTent:2002az} (see also \cite{McAllister:2012am}).

\subsubsection{Quantization}

To discuss the quantization of the two-field system it is convenient to introduce
the canonically normalized fields $q_\varphi \equiv a \varphi$ and $q_\sigma \equiv a \sigma$.
We combine these fields and their conjugate momenta into two-component vectors
\beq
q =  \left( \begin{array}{c}  q_\varphi \\  q_\sigma \end{array} \right) \qquad \Rightarrow \qquad \pi \equiv \frac{\partial {\cal L}}{\partial q'} = \left( \begin{array}{c}  q_\varphi' \\  q_\sigma' \end{array} \right)\ .
\eeq
The fields mix via the following equations of motion
\begin{align}
q_\varphi'' + \left(k^2-\frac{2}{\tau^2}\right)q_\varphi &= +\frac{\rho}{\tau}\left(q_\sigma'-\frac{2}{\tau}q_\sigma\right)\ , \label{equ:EOM1}\\
q_\sigma'' + \left(k^2+ \frac{m^2-2}{\tau^2}\right)q_\sigma &= - \frac{\rho}{\tau}\left(q_\varphi'+\frac{1}{\tau}q_\varphi\right)\ . \label{equ:EOM2}
\end{align}
In order to take into account the statistical independence of the two solutions we have to solve the equations {\it twice}.  In the first run, the inflaton fluctuation~$q_\varphi$ starts in the Bunch-Davies (BD) vacuum, while the second field $q_\sigma$ is set to zero.\footnote{Actually, this initial condition is only consistent in the limit $\rho \to 0$. For finite $\rho$, the equations of motion dictate a correction to the initial condition of $q_\sigma$. We treat this carefully below.}
We denote the corresponding solutions $q^{(\varphi)} = (q_\varphi^{(\varphi)}\, , q_\sigma^{(\varphi)})^T$.  In the second run, $q_\sigma$ starts in Bunch-Davies and $q_\varphi$ is set to zero. The corresponding mode functions are $q^{(\sigma)}$. We can combine the solutions from the two runs into a matrix with components $Q_{I(J)} \equiv q_I^{(J)}$, with $I,J \in \{\varphi, \sigma\}$, i.e.
\beq
Q = \left( \begin{array}{cc} q_\varphi^{(\varphi)} & q_\varphi^{(\sigma)} \\
q_\sigma^{(\varphi)} & q_\sigma^{(\sigma)}
 \end{array} \right) \ .
\eeq
The field operators are then written as\footnote{In component form, this equation reads $\hat q_I = Q_{I(J)} \hat a^{(J)} +  (h.c.)$.}
\beq
\hat q_{\k}(\tau) = Q(k,\tau)\, \hat a_{\k} \ + \ h.c. \ , \qquad {\rm where} \qquad \hat a_{\k} \equiv \big(  \hat a_{\k}^{(\varphi)}\, , \hat a_{\k}^{(\sigma)} \big)^T \ .
\eeq
Imposing the following normalization conditions~\cite{vanTent:2002az},
\begin{align}
Q Q^{*T} - Q^* Q^T &= 0 \ ,\\  \label{unitary}
\Pi \, \Pi^{*T} - \Pi^* \Pi^T &= 0 \ , \\
 Q \Pi^{*T} - Q^* \Pi^{T} &= i \mathbf{1}\label{eq:canonical}\ ,
\end{align}
we get
\beq
\Big[ \hat a_{\k}^{(I)}, (\hat a_{\k'}^{(J)})^\dagger\Big] = \delta^{IJ} \delta(\k - \k')\  ,
\eeq
so that $\hat a_{\k}^{(I)}$ and $(\hat a_{\k'}^{(J)})^\dagger$ are a set of independent annihilation and creation operators, respectively.

\subsubsection{Initial Conditions}

To determine the initial conditions for the coupled two-field system, we consider a WKB solution to the early time limit of the equations of motion, i.e.~when $\omega \gg \rho$.

\vskip 4pt
\noindent
{\it WKB solutions.}---At early times, we expect $\omega \to k + {\cal O}(\tau^{-1})$ and hence we have $\omega' \to {\cal O}(\tau^{-2})$. We wish to solve the equations up to terms that are suppressed by $\tau^{-2}$ (i.e.~we will keep terms of order~$\tau^{-1}$).  The WKB ansatz is
\begin{align}
q_\varphi &= \frac{q_{\varphi,0}}{\sqrt{\omega(\tau)}}\, e^{-i \int^\tau \d \tau' \omega(\tau')}\ , \\
q_\sigma &= \frac{q_{\sigma,0}}{\sqrt{\omega(\tau)}}e^{-i \int^\tau \d \tau' \omega(\tau')}\ .
\end{align}
Plugging these into the equations of motion (\ref{equ:EOM1}) and (\ref{equ:EOM2}), and dropping terms of order $\tau^{-2}$ or higher, we find
\begin{align}
\left(-\omega^2 + k^2\right)q_{\varphi,0}  +i\frac{\rho}{\tau}\omega \hskip 1pt q_{\sigma,0} &=0\ , \label{equ:24}\\
\left(-\omega^2+ k^2 \right)q_{\sigma,0} -i\frac{\rho}{\tau}\omega\hskip 1pt q_{\varphi,0} & =0\ .
\end{align}
Combining these two equations, we find
\beq
\omega^2 - k^2=\  \pm\hskip 1pt  \frac{\rho}{\tau} \omega\ .
 \eeq
 For each sign, this has two solutions corresponding to the positive and negative frequency solutions at early times. We focus on the positive frequency solutions 
 \beq
 \omega_\pm = k\pm\frac{\rho}{2\tau}\ .  \label{equ:25}
 \eeq
  Substituting (\ref{equ:25}) into (\ref{equ:24}), we get a relation between $q_{\varphi,0}$ and $q_{\sigma,0}$,
\beq
q_{\varphi,0}^{\mathsmaller{(\pm)}} = \pm \hskip 1pt i \hskip 1pt q_{\sigma,0}^{\mathsmaller{(\pm)}}\ ,
\eeq
where $\big(q_{\varphi,0}^{\mathsmaller{(\pm)}} ,q_{\sigma,0}^{\mathsmaller{(\pm)}} \big)$ are the amplitudes corresponding to $\omega=\omega_{\pm}$.
We see that $\rho$ has cancelled out, so that $q_{\varphi,0}$ and $q_{\sigma,0}$ are of the same order.
The WKB solutions therefore are 
\begin{align}
q_\varphi^{\mathsmaller{(\pm)}} &= \frac{\pm i}{\sqrt{4 k}} \hskip 1pt e^{-i k\tau} e^{\mp i \frac{\rho}{2} \ln(-k\tau)} \ , \label{equ:q1}\\
q_\sigma^{\mathsmaller{(\pm)}} &= \frac{1}{\sqrt{4 k}} \hskip 1pt e^{-i k\tau} e^{\mp i \frac{\rho}{2} \ln(-k\tau)} \ , \label{equ:q2}
\end{align}
where the overall normalization has been fixed by eq.~(\ref{eq:canonical}).
To make contact with the discussion above, we define alternative basis functions
\begin{align}
q^{(\varphi)} &\equiv \frac{q^{\mathsmaller{(+)}} -q^{\mathsmaller{(-)}}}{i\sqrt{2}} \ , \\
q^{(\sigma)} &\equiv \frac{q^{\mathsmaller{(+)}} +q^{\mathsmaller{(-)}}}{\sqrt{2}} \ .
\end{align}
Using (\ref{equ:q1}) and (\ref{equ:q2}), we find 
\beq
Q = \left( \begin{array}{cc} q_\varphi^{(\varphi)} & q_\varphi^{(\sigma)} \\
q_\sigma^{(\varphi)} & q_\sigma^{(\sigma)}
 \end{array} \right) = \frac{1}{\sqrt{2k}} \hskip 1pt e^{-ik\tau}  \left( \begin{array}{cc} \cos(\tfrac{\rho}{2} \ln(-k\tau)) & -\sin(\tfrac{\rho}{2} \ln(-k\tau)) \\
\sin(\tfrac{\rho}{2} \ln(-k\tau)) & \ \ \, \cos(\tfrac{\rho}{2} \ln(-k\tau))
 \end{array} \right) \  \xrightarrow{\rho \to 0}\ \frac{1}{\sqrt{2k}} \hskip 1pt e^{-ik\tau}  \left( \begin{array}{cc} 1 & 0 \\
0 & 1
 \end{array} \right) \ .
\eeq
These are the initial conditions we advertised above.
They are equivalent to the initial conditions defined in the $(\pm)$ basis.
In particular, the Bunch-Davies vacuum---defined as the state annihilated by both $\hat a^{(\varphi)}$ and $\hat a^{(\sigma)}$---is also annihilated by $\hat a^{\mathsmaller{(\pm)}}$.
In the following, we will find it slightly more convenient to work in the $(\pm)$ basis.

\subsubsection{Mode Functions}

This time we have to solve the exact eqs.~(\ref{equ:A2}) and (\ref{equ:A3}).
Let us write
\beq
\varphi_k(\tau) = \frac{1}{k^{3/2}} \cdot \tilde \varphi(u) \qquad {\rm and} \qquad \sigma_k(\tau) =\frac{1}{k^{3/2}} \cdot \tilde \sigma(u)\ ,
\eeq where $u \equiv k \tau$. The equations of motion in a de Sitter background are
\begin{align}
\tilde \varphi'' - \frac{2}{u}   \tilde \varphi' + \tilde \varphi &\ =\ + \frac{\rho}{u} \left[  \tilde \sigma' - \frac{3}{u} \tilde  \sigma \right] \ , \label{equ:A2x}\\
 \tilde \sigma'' - \frac{2}{u} \tilde \sigma' + \left[1+ \frac{m^2}{u^2} \right] \tilde \sigma  &\ =\ - \frac{\rho}{u}  \hskip 1pt \tilde \varphi'\ . \label{equ:A3x}
\end{align}

\vskip 4pt
\noindent
{\it Initial conditions.}---The two numerical runs are defined by the following set of initial conditions
\begin{align}
&(+)\, : \qquad \lim_{u \to -\infty} \tilde \varphi^{\mathsmaller{(+)}} = i\frac{u}{2}  \hskip 1pt e^{-i(u+\frac{\rho}{2}\ln(-u))}\ , \qquad \lim_{u \to -\infty} \tilde \sigma^{\mathsmaller{(+)}} =   \frac{u}{2}  \hskip 1pt e^{-i(u+\frac{\rho}{2}\ln(-u))}  \ , \\
&(-)\, : \qquad \lim_{u \to -\infty} \tilde \varphi^{\mathsmaller{(-)}} =   -i\frac{u}{2}  \hskip 1pt e^{-i(u-\frac{\rho}{2}\ln(-u))} \ , \qquad \lim_{u \to -\infty}   \tilde \sigma^{\mathsmaller{(-)}} = \frac{u}{2} \hskip 1pt e^{-i (u-\frac{\rho}{2}\ln(-u))} \ .
\end{align}

\vskip 4pt
\noindent
{\it Wick-rotated solutions.}---This time we perform the Wick rotation $u \to i u$. We define Wick-rotated mode functions as
\beq
\tilde \varphi_w(u) \equiv \tilde \varphi(iu)\quad{\rm and}\quad \tilde \sigma_w(u) \equiv \tilde \sigma(iu).
\eeq
Eqs.~(\ref{equ:A2x}) and (\ref{equ:A3x}) become
\begin{align}
\tilde \varphi''_w - \frac{2}{u}   \tilde \varphi'_w - \tilde \varphi_w &\ =\ + \frac{\rho}{u} \left[  \tilde \sigma'_w - \frac{3}{u} \tilde  \sigma_w \right] \ , \\
 \tilde \sigma''_w - \frac{2}{u} \tilde \sigma'_w - \left[1- \frac{m^2}{u^2} \right] \tilde \sigma_w  &\ =\ - \frac{\rho}{u}  \hskip 1pt \tilde \varphi_w'\ ,
\end{align}
with initial conditions
\begin{align}
&(+)\, : \qquad \lim_{u \to -\infty} \tilde \varphi^{\mathsmaller{(+)}}_w =  i\frac{u}{2}  \hskip 1pt e^{u-i\frac{\rho}{2}\ln(-u)}e^{\frac{\pi}{4}\rho}\ , \qquad \lim_{u \to -\infty} \tilde \sigma_w^{\mathsmaller{(+)}} =   \frac{u}{2}  \hskip 1pt e^{u-i\frac{\rho}{2}\ln(-u)}e^{\frac{\pi}{4}\rho}  \ , \\
&(-)\, : \qquad \lim_{u \to -\infty} \tilde \varphi^{\mathsmaller{(-)}}_w =  -i \frac{u}{2}  \hskip 1pt e^{u+i\frac{\rho}{2}\ln(-u)}e^{-\frac{\pi}{4}\rho} \ , \qquad \lim_{u \to -\infty}   \tilde \sigma^{\mathsmaller{(-)}}_w = \frac{u}{2} \hskip 1pt e^{u+i\frac{\rho}{2}\ln(-u)}e^{-\frac{\pi}{4}\rho} \ .
\end{align}

\subsubsection{Power Spectrum}

The power spectrum of curvature perturbations is an incoherent superposition of the results of the two runs,
\beq
(2\pi)^2 \Delta_\zeta^2 = \frac{2}{\dot \Phi^2_0} \left[ \big|\tilde \varphi^{\mathsmaller{(+)}}(0)\big|^2 +  \big|\tilde \varphi^{\mathsmaller{(-)}}(0) \big|^2 \right]\ .
\eeq

\subsubsection{Bispectra}

We define the following momentum-dependent mode functions:
\beq
\varphi_{w,k}(v)\equiv \frac{1}{k^{3/2}}\cdot\tilde\varphi_{w}(k v)\ .
\eeq
Wick contractions between the fields $\varphi$ and $\sigma$ are
\begin{align}
\contraction{}{\hat\varphi}{_{w,k}(v_{1})}{\hat\varphi}\hat\varphi_{w,k} (v_{1})\hat\varphi_{w,k}( v_{2})
&\equiv \langle0|\hat\varphi_{w,k} (v_{1})\hat\varphi_{w,k}(v_{2})|0\rangle
 = \sum_{\alpha=\pm}\varphi^{{(\alpha)}}_{w,k}(v_{1})\big[\varphi^{{(\alpha)}}_{w,k}( v_{2})\big]^*\ ,\\
\contraction{}{\hat\varphi}{_{w,k}(v_{1})}{\hat\sigma}\hat\varphi_{w,k} (v_{1})\hat\sigma_{w,k}(v_{2})&\equiv \langle0|\hat\varphi_{w,k} (v_{1})\hat\sigma_{w,k}( v_{2})|0\rangle = \sum_{\alpha=\pm}\varphi^{{(\alpha)}}_{w,k}(v_{1})\big[\sigma^{{(\alpha)}}_{w,k}(v_{2})\big]^*\ .
\end{align}

\vskip 4pt

\noindent{\it Bispectrum from $(\partial\varphi)^2\sigma$.---}The interaction $H_{\rm int}^{\mathsmaller{(1)}}=\frac{1}{2}\frac{(\partial\varphi)^2\sigma}{\Lambda}$ gives the bispectrum
\begin{align}
B_\varphi^{\mathsmaller{(1)}}(k_1,k_2,k_3) & = \frac{1}{\Lambda} \hskip 1pt {\cal I}^{\mathsmaller{(1)}}(k_1,k_2,k_3)\ ,
\end{align}
where ${\cal I}^{\mathsmaller{(1)}}$ is the sum of two terms, ${\cal I}_{\mathsmaller{\dot\varphi^2\sigma}}$ and ${\cal I}_{\mathsmaller{(\partial_i\varphi)^2\sigma}} $, which arise from the interactions generated by the $\dot\varphi^2\sigma$ and $(\partial_i\varphi)^2\sigma$ terms, respectively.  These two terms read
\begin{align}
{\cal I}_{\mathsmaller{\dot\varphi^2\sigma}}(k_1,k_2,k_3) &\ =\ 2{\rm Re}\bigg[\int_{-\infty}^0\frac{\d v}{v^2}\big[\contraction{}{\hat\varphi}{_{w,k_1}(0)}{\hat\varphi}\hat\varphi_{w,k_1} (0)\hat\varphi_{w,k_1}'(v)\big] \, \nonumber\\
&  \hskip 80pt\times\big[\contraction{}{\hat\varphi}{_{w,k_2}(0)}{\hat\varphi}\hat\varphi_{w,k_2} (0)\hat\varphi_{w,k_2}'(v)\big]\big[\contraction{}{\hat\varphi}{_{w,k_3}(0)}{\hat\sigma}\hat\varphi_{w,k_3} (0)\hat\sigma_{w,k_3}(v)\big] +2\,{\rm perms.} \bigg]\ ,\\
{\cal I}_{\mathsmaller{(\partial_i\varphi)^2\sigma}}(k_1,k_2,k_3) &\ =\ -{\rm Re}\bigg[\int_{-\infty}^0\frac{\d v}{v^2} \, \big(k_3^2-k_2^2-k_1^2\big)\nonumber\\
&  \hskip 40pt \times \big[\contraction{}{\hat\varphi}{_{w,k_1}(0)}{\hat\varphi}\hat\varphi_{w,k_1} (0)\hat\varphi_{w,k_1}(v)\big]\big[\contraction{}{\hat\varphi}{_{w,k_2}(0)}{\hat\varphi}\hat\varphi_{w,k_2} (0)\hat\varphi_{w,k_2}(v)\big]\big[\contraction{}{\hat\varphi}{_{w,k_3}(0)}{\hat\sigma}\hat\varphi_{w,k_3} (0)\hat\sigma_{w,k_3}(v)\big] +2\,{\rm perms.} \bigg]\ .
\end{align}


\vskip 4pt
\noindent{\it Bispectrum from $\sigma^3$.---}The interaction $H_{\rm int}^{\mathsmaller{(2)}}=\mu\sigma^3$ gives
\begin{align}
B_\varphi^{\mathsmaller{(2)}}(k_1,k_2,k_3) &= \mu \hskip 2pt {\cal I}^{\mathsmaller{(2)}}(k_1,k_2,k_3)\ ,
\end{align}
with
\begin{align}
{\cal I}^{\mathsmaller{(2)}}(k_1,k_2,k_3) &\ =\ -12\hskip 1pt \mu\,{\rm Re}\bigg[\int_{-\infty}^0\frac{\d v}{v^4} \big[\contraction{}{\hat\varphi}{_{w,k_1}(0)}{\hat\sigma}\hat\varphi_{w,k_1} (0)\hat\sigma_{w,k_1}(v)\big]\big[\contraction{}{\hat\varphi}{_{w,k_2}(0)}{\hat\sigma}\hat\varphi_{w,k_2} (0)\hat\sigma_{w,k_2}(v)\big]\big[\contraction{}{\hat\varphi}{_{w,k_3}(0)}{\hat\sigma}\hat\varphi_{w,k_3} (0)\hat\sigma_{w,k_3}(v)\big]\bigg]\ .
\end{align}
 
\noindent{\it Full bispectrum.}---The full bispectrum of the primordial curvature perturbations is
\begin{align}
B_\zeta(k_1,k_2,k_3) &\ =\ -\left( \frac{H}{\dot \Phi_0}\right)^3 \bigg[ B_\varphi^{\mathsmaller{(1)}} + B_\varphi^{\mathsmaller{(2)}} \bigg] \ .
\end{align}
Using eq.~(\ref{equ:fnlDEF}), we get
\beq
\fnl= f_1(m,\rho) \cdot \frac{\rho}{H} +f_2(m,\rho) \cdot \frac{1}{2\pi \Delta_\zeta} \cdot \frac{\mu}{H} \ ,
\eeq
where
\beq
f_1(m,\rho) \equiv -\frac{5}{18}\,\frac{{\cal I}^{\mathsmaller{(1)}}(k,k,k)}{P^2_\varphi(k)}\qquad {\rm and} \qquad
f_2(m,\rho) \equiv  -2\pi\Delta_\varphi\cdot\frac{5}{18}\,\frac{{\cal I}^{\mathsmaller{(2)}}(k,k,k)}{P^2_\varphi(k)}\ .
\eeq

\begin{figure}[h!]
   \centering
       \includegraphics[scale =0.35]{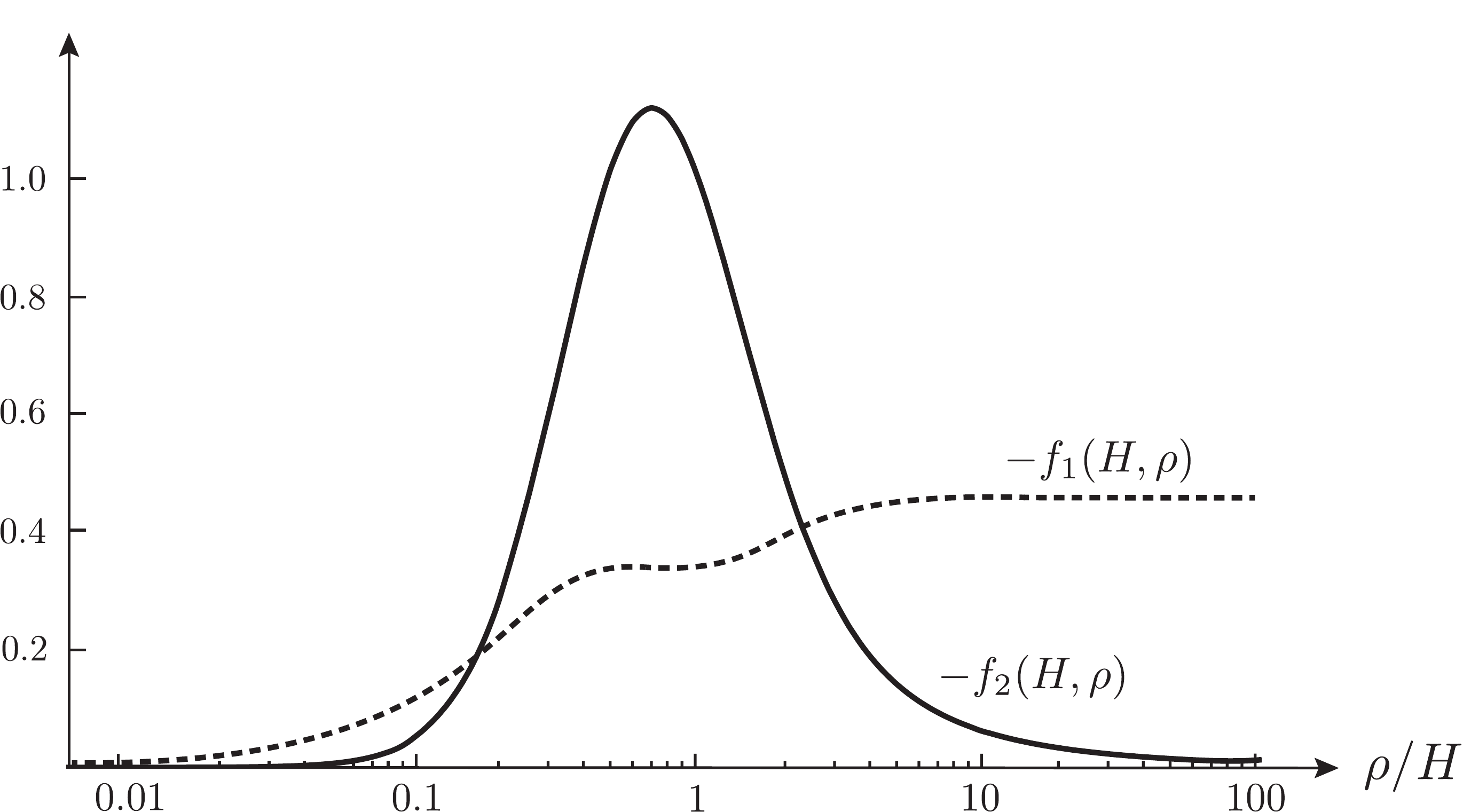}
   \caption{Numerical computation of $f_1(m,\rho)$ (dashed) and $f_2(m,\rho)$ (solid) for $m=H$.}
  \label{fig:functions2}
\end{figure}

\noindent{\it IR divergences.---}The factorized form of the {\it in-in} formula is not well-suited to deal with the IR behavior of the integral, i.e.~the behavior as $v\to0$. This leads to a spurious IR divergence when the integral is computed numerically. In order to make the IR convergence manifest, one needs to use the so-called commutator form, which schematically reads
\beq
\langle\hat \varphi^3\rangle = \int^0_{-\infty}\d\tau\,\langle[\hat H_{\rm int},\hat\varphi^3]\rangle\ .
\eeq
However, the Wick rotation is no longer valid in the commutator form and therefore one cannot achieve convergence in the UV. Hence, one way to make the integral converge both in the UV and the IR is to split it in two parts~\cite{Chen:2009zp}
\beq
\langle\hat \varphi^3\rangle =2{\rm Re}\Bigg[-i\int_{-\infty(1-i\epsilon)}^{\tau_c}\langle\hat\varphi^3H_{\rm int}(\tau)\rangle\Bigg] +\int_{\tau_c}^0\d\tau\langle[\hat H_{\rm int}(\tau),\hat\varphi^3]\rangle\ ,
\eeq
where the UV part ($\tau<\tau_c$) is computed in the factorized form and then Wick rotated, while the IR part ($\tau>\tau_c$) is computed in the commutator form. This solves the problem of spurious IR divergences. Naturally, the final result is independent of the choice of cutoff $\tau_c$ used to distinguish the UV and IR parts of the integral.

\newpage
\section{Details of the Effective Theory}
\label{app:EFT}

The operator (\ref{equ:Lmix}) that mixes the inflaton sector $\Phi$ and the hidden sector $\Sigma$ contains a tadpole for~$\Sigma$,
\beq
{\cal L}_{\rm mix} \supset (\rho \dot \Phi_0) \hskip 1pt \Sigma \ .
\eeq
In Section~\ref{sec:EFT}, we assumed that a potential $V(\Sigma)$ stabilizes the field at $\Sigma_0$ and then studied the phenomenology of fluctuations $\sigma \equiv \Sigma - \Sigma_0$.
In this appendix, we will discuss this problem in more detail.
 Because $\dot \Phi_0 \gg H^2$, we expect the minimum $\Sigma_0$ to be displaced significantly from the origin.
The effective couplings of fluctuations around $\Sigma_0$ can therefore be quite different from those near the origin. In the cases of greatest phenomenological interest, all dimensionful couplings in the effective action for $\sigma$ were of order the Hubble scale $H$. Here,
we want to explore whether this structure is natural  after including the vev for $\Sigma$.
We will describe two different vantage points: the effective theory of the background~\cite{Weinberg:2008hq} and the effective theory of fluctuations~\cite{Cheung:2007st}. For concreteness, we will limit the discussion to the weak mixing regime.

\subsection{Effective Theory of the Background}

We start with a simple model in which a single scale controls both the mixing term and the hidden sector self-interactions.  
For weak mixing, this model will fail to produce measurable non-Gaussianity, but the way in which it fails will be instructive.
In particular, it will motivate models in which the mixing interaction and the hidden sector self-interactions are controlled by two distinct scales.

\vskip 4pt
\noindent
{\it One-scale models.}---Consider the canonical slow-roll Lagrangian
\beq
{\cal L}_{\Phi} = - \frac{1}{2}(\partial \Phi)^2 - V(\Phi)\ .
\eeq
We include a tree-level coupling to a free field $\Sigma$,
\beq\label{equ:onescaleL}
\Big[{\cal L}_\Sigma + {\cal L}_{\rm mix}\Big]_{\rm tree} = -\frac{1}{2} (\partial \Sigma )^2 -  \frac{1}{2} \frac{(\partial \Phi )^2}{\Lambda} \Sigma \ .
\eeq
We have chosen  the classical potential for $\Sigma$ to vanish, so that  $\Sigma$ is protected by a shift symmetry that is broken only by the higher-dimensional mixing term.  All corrections to the potential for $\Sigma$ therefore vanish as $\Lambda \to \infty$.
Let us assume that this effective description holds up to a cutoff scale~$\Lambda_\star$.  The one-loop effective action  for $\Sigma$ is then
\beq
\Big[ {\cal L}_\Sigma + {\cal L}_{\rm mix} \Big]_{\text{1-loop}} = -\frac{1}{2} (\partial \Sigma)^2 -  \frac{1}{2}\frac{(\partial \Phi )^2}{\Lambda} \Sigma - \frac{\Lambda_{\star}^4}{16 \pi^2 \Lambda^2} \, \Sigma^2 \left[ 1 + c_1 \frac{\Sigma}{ \Lambda} + c_2 \frac{\Sigma^2}{\Lambda^2} + \cdots \right]  \ . \label{equ:Lloop}
\eeq
We have cancelled the one-loop tadpole, since we are assuming that $\langle \Sigma\rangle=0$ in the vacuum.  In order to induced a mass of at most $m^2 $, we require 
that $\Lambda^4_{\star} \lesssim 16 \pi^2 m^2 \Lambda^2 $.
Using this mass term, $m^2 \Sigma^2$, to stabilize the tadpole for $\Sigma$, we find
\beq
\Sigma_0 = \frac{\dot \Phi_0^2}{\Lambda m^2} = \left( \frac{\rho}{m} \right)^2 \, \Lambda \ . \label{equ:S0}
\eeq
For strong mixing, we require $\rho \gg M \equiv m^2 / \rho$, which implies that $\Sigma_0 \gg \Lambda$.  As a result, in the case of strong mixing, the EFT description breaks down between $\Sigma=0$ and $\Sigma = \Sigma_0$.  We will discuss the EFT around $\Sigma_0$ in the next subsection.

The case of weak mixing ($\rho \lesssim H$) is much less constrained: we are free to take $\rho \lesssim m \lesssim H$, so that $\Sigma_0 < \Lambda$.
When this holds, higher-order terms in the potential are suppressed by $\Sigma_0 / \Lambda$ and are therefore not important.
The structure of this effective theory is appealing: the small mass for $\Sigma$ is explained by the approximate shift symmetry for $\Sigma$ (assuming an appropriate UV completion above $\Lambda_\star$).  Furthermore, we have found that the vev, $\Sigma_0$, is under control: higher-dimension operators are suppressed by powers of $\Sigma_0 / \Lambda$.

Unfortunately, for weak mixing this setup is too restrictive to allow for measurable non-Gaussianity.  To see this we note that the  cubic coupling in (\ref{equ:Lloop}) is $\mu \sim H^2/ \Lambda$.
Eq.~(\ref{equ:fQSFI}) then implies
\beq
\fnl = f(m) \cdot \frac{1}{2\pi \Delta_\zeta} \cdot \frac{H}{\Lambda} \left( \frac{\rho}{H}\right)^{3} = f(m) \left( \frac{\rho}{H} \right)^4 < \frac{f(m)}{c^2(m)} \lesssim  1\ , \label{equ:f0}
\eeq
where we have used $(\rho/H)^2 < c^{-1}(m)$ as required for perturbative control of the power spectrum~(\ref{equ:PS}).
We see that the amplitude is parametrically suppressed and unobservable in present and future CMB experiments.
Including a tree-level potential $V(\Sigma) = \mu \Sigma^3$, with $\mu \sim H$, does not improve the situation.  At $\Sigma_0$, this would induce a mass term $m^2 \sim H \Sigma_0$. To keep the field light, we have to require $\Sigma_0 \lesssim H$ (or accept fine-tuning against the bare mass.).
But eq.~(\ref{equ:S0}) then implies $\rho < (2\pi \Delta_\zeta) H$ and we get
\beq
 \fnl < (2\pi \Delta_\zeta)^2 f(m) \sim 10^{-9} f(m)  \ll 1\ .
\eeq
 We conclude that the one-scale model, although natural, is too weakly coupled to produce a significant three-point function.   The same applies to the four-point function,
 \beq
 \gnl = g(m) \cdot \frac{1}{(2\pi \Delta_\zeta)^2}\cdot \left( \frac{H}{\Lambda} \right)^2 \left( \frac{\rho}{H}\right)^{4} = g(m) \left( \frac{\rho}{H} \right)^6 < \frac{g(m)}{c^3(m)} \lesssim 1\ ,  \label{equ:gnlA}
 \eeq
 where we have used the quartic coupling in eq.~(\ref{equ:Lloop}).
 Although we have addressed the small size of the mass term and cancelled the tadpole, we have also suppressed the interactions that generate the non-Gaussian correlations.

\vskip 4pt
\noindent
{\it Two-scale models.}---For weak mixing, in order to produce measurable non-Gaussianity, we need an additional cubic interaction beyond the one induced by the mixing term.  The basic problem we just ran into can be traced back to the fact that we associated the same scale to the breaking of the shift symmetry and to the strength of the cubic (or quartic) interaction.  The obvious strategy to produce a measurable signal is to separate these two scales. The examples that we will present are meant as simple existence proofs of natural theories with large non-Gaussianities, rather than as a comprehensive study of the range of possibilities.

\vskip 4pt
The most straightforward modification to the effective Lagrangian in (\ref{equ:onescaleL}) is to introduce additional interaction terms that do not violate the shift symmetry for $\Sigma$.  These interaction terms can then be large without any risk of introducing large corrections to the mass of $\Sigma$.  As a concrete example, we will consider
\beq
{\cal L}_{\rm \Sigma} \supset \frac{(\partial_\mu \Sigma \partial^\mu \Sigma)^2 }{\tilde \Lambda^4} \ .
\eeq
After including the one-loop corrections~(\ref{equ:Lloop}), the tadpole is stabilized as before.
At $\Sigma_0$, we have the quartic interaction
\beq
{\cal L}_{\rm int} = \frac{1}{\tilde \Lambda^4} (\partial \sigma )^4\ .
\eeq
The associated trispectrum amplitude is
 \beq
 \gnl = \tilde g(m) \cdot \frac{1}{(2\pi \Delta_\zeta)^2}\cdot \left( \frac{H}{\tilde \Lambda} \right)^4 \left( \frac{\rho}{H}\right)^{4} \ .
 \eeq
For $\tilde \Lambda \ll \Lambda$, this can be significantly boosted relative to (\ref{equ:gnlA}).
 An observable trispectrum can arise if $ \tilde \Lambda \lesssim (\dot \Phi_0)^{1/2}$.
The scale $\tilde \Lambda$ is not consistent with the effective theory for the background at a scale of order $(\dot \Phi_0)^{1/2}$.  Nevertheless, couplings of this type are allowed within the effective theory for the fluctuations, provided that $\tilde \Lambda > H = 0.02 \hskip 1pt (\dot \Phi_0)^{1/2}$.  Furthermore, the dynamics that lead to this large interaction are confined to the $\Sigma$ sector and do not affect the evolution of $\Phi_0(t)$.

The above model illustrates how to introduce additional interactions for $\Sigma$ without affecting the mass of the field.  In principle,  the same logic can be used to produce a large bispectrum.  A simple operator that achieves this is
\beq
\frac{1}{\tilde \Lambda^{4}} \partial_\mu \Phi \partial^\mu \Sigma (\partial \Sigma)^2 \ \to\ \frac{\dot \Phi_0}{\tilde \Lambda^4} \hskip 1pt \dot \sigma (\partial \sigma)^2 \ .
\eeq
For small enough $\tilde \Lambda$, this allows measurable non-Gaussianity.
The operator also contains additional mixing terms that transfer the fluctuations to the visible sector.
It would be interesting to explore these model-building considerations further, but this is beyond the scope of the present paper.
 
\subsection{Effective Theory of the Fluctuations}

In expanding $V(\Sigma)$ around $\Sigma_0$, we have assumed a single EFT for the background that is valid around both $\Sigma=0$ and $\Sigma = \Sigma_0$.  Imposing such a constraint is unnecessarily strong, as the $\sigma$ field only probes a small region around $\Sigma_0$.  By contrast, in the effective theory of inflation~\cite{Cheung:2007st}, one writes down the theory for the fluctuations directly.  The only remnants of the background solution are the (time-dependent) couplings in the action.  Specifically, the Lagrangian for the fluctuations is given by
\beq\label{equ:eftL1}
{\cal L} = -\frac{1}{2}(\partial \varphi)^2 - \frac{1}{2} (\partial \sigma)^2 - \frac{1}{\Lambda} \left[  \dot \Phi_0 \dot \varphi - \frac{1}{2} (\partial \varphi)^2 \right] \sigma - V(\sigma) \ .
\eeq
By writing the mixing term as $[\dot \Phi_0 \dot \varphi - \tfrac{1}{2} (\partial \varphi)^2]\sigma $, we have explicitly\footnote{This is more transparent in the language of the EFT of inflation~\cite{Cheung:2007st}, where this mixing term is written as ${\cal L}_{\rm mix} = -\dot \Phi_0^2 [ \partial (t+\pi)^2 + 1] \sigma$ with $\pi \equiv \varphi / \dot \Phi$.  The tadpole had to be cancelled by hand by including the $+1$, otherwise the action would break diffeomorphism invariance explicitly.}   removed the tadpole for $\sigma$, which is stabilized at $\sigma = 0$ for all time.  As in the main text, we take $V(\sigma) = \tfrac{1}{2} m^2 \sigma^2 + \mu \sigma^3 + \cdots$.  It is the parameters $\Lambda$, $m^2$ and $\mu$ that are constrained by Planck's limits on non-Gaussianity.  

In the main text, we have shown that the mass and the cubic coupling receive the one-loop corrections
\beq
\delta m^2 \sim \frac{\Lambda_\star^4}{\Lambda^2} \qquad {\rm and} \qquad \delta\mu \sim \frac{\Lambda_\star^4}{\Lambda^3}\ .
\eeq
In the case of weak mixing, $m^2 \lesssim H^2 $ and $\rho \ , \, \mu   \lesssim H$, these are small corrections when $\Lambda_\star \sim (\dot \Phi_0)^{1/2}$.  For strong mixing, $\rho \gg H$, these corrections are more dangerous but still produce a natural EFT when $\Lambda_\star \sim \rho < (\dot \Phi_0)^{1/2}$.
One may be tempted to conclude that there is no fine tuning in the EFT of the fluctuations.
On the other hand, this does not seem to match our experience in the case of slow-roll inflation with a polynomial potential for $\Sigma$.  The resolution is that, for slow-roll inflation, the interactions in the effective Lagrangian are negligible and we are free to take $\Lambda_{\star} \sim \Lambda$.  Under such circumstances, we would find $\delta m^2 \sim \Lambda^2 \gg \dot \Phi_0$.  We conclude that without a new scale below $\Lambda$, writing a small mass for $\sigma$ is not natural (for weak or strong mixing).

\subsection{Summary}

In the previous two subsections, we described the naturalness of the EFT from the point of view of the background and the fluctuations.  In terms of the background, we found that our EFT around $\Sigma= 0$ can break down before we reach $\Sigma = \Sigma_0$.  On the other hand, the theory of the fluctuations is well-defined and natural, provided that the theory is cut off at $\Lambda_\star \lesssim (\dot \Phi_0)^{1/2}$.

To make contact between the two descriptions, let us consider an effective theory for the background that is well-defined for a region of size $\Lambda_\star < \Lambda$ around $\Sigma_0$.  For example, near $\Sigma_0$ we might have
\beq
V(\Sigma) = M^{4-\alpha} \Sigma^\alpha + m_0^2 (\Sigma-\Sigma_0)^2 +  \mu_0 (\Sigma - \Sigma_0)^3 + \cdots \ ,
\eeq
 where $1<\alpha <2$.  The first term in the potential is responsible for canceling the tadpole at $\Sigma = \Sigma_0$.  Expanding the leading term around $\Sigma = \Sigma_0$, we find that the corrections to the mass and to the cubic coupling are given by
 \beq
 \delta m^2 \sim   \frac{\rho\hskip 1pt \dot \Phi_0} { \Sigma_0} \qquad {\rm and} \qquad \delta \mu \sim   \frac{\rho \hskip 1pt \dot \Phi_0} { \Sigma_0^2}\ .
 \eeq
 Taking $\Sigma_0 \gtrsim \dot \Phi_0 / H$ ensures that these corrections are negligible.  However, for the loop corrections to be small, our ``effective theory" must be cut off at $\Lambda_{\star} \sim (\dot \Phi)^{1/2}_0 \ll \Sigma_0$.  Therefore, this effective theory does not include $\Sigma=0$.

 \vskip 4pt
In conclusion, there exist effective theories for the fluctuations where the mass for $\sigma$ is naturally small, yet there is still a large non-Gaussian signal.  However, a single weakly coupled effective theory does not include both the vacuum and inflationary values of $\Sigma$.

\newpage
\addcontentsline{toc}{section}{References}

\bibliographystyle{utphys}
\bibliography{HighScale-Refs}

\providecommand{\href}[2]{#2}\begingroup\raggedright\begin{thebibliography}{10}

\bibitem{PlanckParameters}
P.~Ade {\em et~al.}, ``{Planck 2013 Results. XVI. Cosmological Parameters},''
\href{http://arxiv.org/abs/1303.5076}{{\ttfamily arXiv:1303.5076
  [astro-ph.CO]}}.

\bibitem{PlanckInflation}
P.~Ade {\em et~al.}, ``{Planck 2013 Results. XXII. Constraints on Inflation},''
\href{http://arxiv.org/abs/1303.5082}{{\ttfamily arXiv:1303.5082
  [astro-ph.CO]}}.

\bibitem{PlanckNG}
P.~Ade {\em et~al.}, ``{Planck 2013 Results. XXIV. Constraints on Primordial
  Non-Gaussianity},''
\href{http://arxiv.org/abs/1303.5084}{{\ttfamily arXiv:1303.5084
  [astro-ph.CO]}}.

\bibitem{Gangui:1993tt}
A.~Gangui, F.~Lucchin, S.~Matarrese, and S.~Mollerach, ``{The Three-Point
  Correlation Function of the Cosmic Microwave Background in Inflationary
  Models},'' \href{http://dx.doi.org/10.1086/174421}{{\em Astrophys.J.}
  {\bfseries 430} (1994) 447--457},
\href{http://arxiv.org/abs/astro-ph/9312033}{{\ttfamily arXiv:astro-ph/9312033
  [astro-ph]}}.

\bibitem{Babich:2004gb}
D.~Babich, P.~Creminelli, and M.~Zaldarriaga, ``{The Shape of
  Non-Gaussianities},''
  \href{http://dx.doi.org/10.1088/1475-7516/2004/08/009}{{\em JCAP} {\bfseries
  0408} (2004) 009},
\href{http://arxiv.org/abs/astro-ph/0405356}{{\ttfamily arXiv:astro-ph/0405356
  [astro-ph]}}.

\bibitem{Senatore:2009gt}
L.~Senatore, K.~M. Smith, and M.~Zaldarriaga, ``{Non-Gaussianities in
  Single-Field Inflation and their Optimal Limits from the WMAP 5-year Data},''
  \href{http://dx.doi.org/10.1088/1475-7516/2010/01/028}{{\em JCAP} {\bfseries
  1001} (2010) 028},
\href{http://arxiv.org/abs/0905.3746}{{\ttfamily arXiv:0905.3746
  [astro-ph.CO]}}.

\bibitem{Cheung:2007st}
C.~Cheung, P.~Creminelli, A.~L. Fitzpatrick, J.~Kaplan, and L.~Senatore, ``{The
  Effective Field Theory of Inflation},''
  \href{http://dx.doi.org/10.1088/1126-6708/2008/03/014}{{\em JHEP} {\bfseries
  0803} (2008) 014},
\href{http://arxiv.org/abs/0709.0293}{{\ttfamily arXiv:0709.0293 [hep-th]}}.

\bibitem{Baumann:2011su}
D.~Baumann and D.~Green, ``{Equilateral Non-Gaussianity and New Physics on the
  Horizon},'' \href{http://dx.doi.org/10.1088/1475-7516/2011/09/014}{{\em JCAP}
  {\bfseries 1109} (2011) 014},
\href{http://arxiv.org/abs/1102.5343}{{\ttfamily arXiv:1102.5343 [hep-th]}}.

\bibitem{Maldacena:2002vr}
J.~M. Maldacena, ``{Non-Gaussian Features of Primordial Fluctuations in
  Single-Field Inflationary Models},'' {\em JHEP} {\bfseries 0305} (2003) 013,
\href{http://arxiv.org/abs/astro-ph/0210603}{{\ttfamily arXiv:astro-ph/0210603
  [astro-ph]}}.

\bibitem{Creminelli:2011mw}
P.~Creminelli, ``{Conformal Invariance of Scalar Perturbations in Inflation},''
  \href{http://dx.doi.org/10.1103/PhysRevD.85.041302}{{\em Phys.Rev.}
  {\bfseries D85} (2012) 041302},
\href{http://arxiv.org/abs/1108.0874}{{\ttfamily arXiv:1108.0874 [hep-th]}}.

\bibitem{BM}
D.~Baumann and L.~McAllister, ``{Inflation and String Theory},'' 2013.

\bibitem{Green:2013rd}
D.~Green, M.~Lewandowski, L.~Senatore, E.~Silverstein, and M.~Zaldarriaga,
  ``{Anomalous Dimensions and Non-Gaussianity},''
\href{http://arxiv.org/abs/1301.2630}{{\ttfamily arXiv:1301.2630 [hep-th]}}.

\bibitem{Skiba:2010xn}
W.~Skiba, ``{TASI Lectures on Effective Field Theory and Precision Electroweak
  Measurements},''
\href{http://arxiv.org/abs/1006.2142}{{\ttfamily arXiv:1006.2142 [hep-ph]}}.

\bibitem{Beringer:1900zz}
J.~Beringer {\em et~al.}, ``{Review of Particle Physics},''
\href{http://dx.doi.org/10.1103/PhysRevD.86.010001}{{\em Phys.Rev.} {\bfseries
  D86} (2012) 010001}.

\bibitem{Baumann:2011nk}
D.~Baumann and D.~Green, ``{Signatures of Supersymmetry from the Early
  Universe},'' \href{http://dx.doi.org/10.1103/PhysRevD.85.103520}{{\em
  Phys.Rev.} {\bfseries D85} (2012) 103520},
\href{http://arxiv.org/abs/1109.0292}{{\ttfamily arXiv:1109.0292 [hep-th]}}.

\bibitem{Senatore:2010wk}
L.~Senatore and M.~Zaldarriaga, ``{The Effective Field Theory of Multi-Field
  Inflation},'' \href{http://dx.doi.org/10.1007/JHEP04(2012)024}{{\em JHEP}
  {\bfseries 1204} (2012) 024},
\href{http://arxiv.org/abs/1009.2093}{{\ttfamily arXiv:1009.2093 [hep-th]}}.

\bibitem{Noumi:2012vr}
T.~Noumi, M.~Yamaguchi, and D.~Yokoyama, ``{Effective Field Theory Approach to
  Quasi-Single-Field Inflation},''
\href{http://arxiv.org/abs/1211.1624}{{\ttfamily arXiv:1211.1624 [hep-th]}}.

\bibitem{Weinberg:2008hq}
S.~Weinberg, ``{Effective Field Theory for Inflation},''
  \href{http://dx.doi.org/10.1103/PhysRevD.77.123541}{{\em Phys.Rev.}
  {\bfseries D77} (2008) 123541},
\href{http://arxiv.org/abs/0804.4291}{{\ttfamily arXiv:0804.4291 [hep-th]}}.

\bibitem{Creminelli:2003iq}
P.~Creminelli, ``{On Non-Gaussianities in Single-Field Inflation},''
  \href{http://dx.doi.org/10.1088/1475-7516/2003/10/003}{{\em JCAP} {\bfseries
  0310} (2003) 003},
\href{http://arxiv.org/abs/astro-ph/0306122}{{\ttfamily arXiv:astro-ph/0306122
  [astro-ph]}}.

\bibitem{Silverstein:2003hf}
E.~Silverstein and D.~Tong, ``{Scalar Speed Limits and Cosmology: Acceleration
  from D-cceleration},''
  \href{http://dx.doi.org/10.1103/PhysRevD.70.103505}{{\em Phys.Rev.}
  {\bfseries D70} (2004) 103505},
\href{http://arxiv.org/abs/hep-th/0310221}{{\ttfamily arXiv:hep-th/0310221
  [hep-th]}}.

\bibitem{Baumann:2009ds}
D.~Baumann, ``{TASI Lectures on Inflation},''
\href{http://arxiv.org/abs/0907.5424}{{\ttfamily arXiv:0907.5424 [hep-th]}}.

\bibitem{Weinberg:2005vy}
S.~Weinberg, ``{Quantum Contributions to Cosmological Correlations},''
  \href{http://dx.doi.org/10.1103/PhysRevD.72.043514}{{\em Phys.Rev.}
  {\bfseries D72} (2005) 043514},
\href{http://arxiv.org/abs/hep-th/0506236}{{\ttfamily arXiv:hep-th/0506236
  [hep-th]}}.

\bibitem{Chen:2009zp}
X.~Chen and Y.~Wang, ``{Quasi-Single-Field Inflation and Non-Gaussianities},''
  \href{http://dx.doi.org/10.1088/1475-7516/2010/04/027}{{\em JCAP} {\bfseries
  1004} (2010) 027},
\href{http://arxiv.org/abs/0911.3380}{{\ttfamily arXiv:0911.3380 [hep-th]}}.

\bibitem{Tolley:2009fg}
A.~J. Tolley and M.~Wyman, ``{The Gelaton Scenario: Equilateral Non-Gaussianity
  from Multi-Field Dynamics},''
  \href{http://dx.doi.org/10.1103/PhysRevD.81.043502}{{\em Phys.Rev.}
  {\bfseries D81} (2010) 043502},
\href{http://arxiv.org/abs/0910.1853}{{\ttfamily arXiv:0910.1853 [hep-th]}}.

\bibitem{Cremonini:2010ua}
S.~Cremonini, Z.~Lalak, and K.~Turzynski, ``{Strongly Coupled Perturbations in
  Two-Field Inflationary Models},''
  \href{http://dx.doi.org/10.1088/1475-7516/2011/03/016}{{\em JCAP} {\bfseries
  1103} (2011) 016},
\href{http://arxiv.org/abs/1010.3021}{{\ttfamily arXiv:1010.3021 [hep-th]}}.

\bibitem{Avgoustidis:2012yc}
A.~Avgoustidis, S.~Cremonini, A.-C. Davis, R.~H. Ribeiro, K.~Turzynski, {\em
  et~al.}, ``{Decoupling Survives Inflation: A Critical Look at Effective Field
  Theory Violations During Inflation},''
  \href{http://dx.doi.org/10.1088/1475-7516/2012/06/025}{{\em JCAP} {\bfseries
  1206} (2012) 025},
\href{http://arxiv.org/abs/1203.0016}{{\ttfamily arXiv:1203.0016 [hep-th]}}.

\bibitem{Achucarro:2010da}
A.~Achucarro, J.-O. Gong, S.~Hardeman, G.~A. Palma, and S.~P. Patil,
  ``{Features of Heavy Physics in the CMB Power Spectrum},''
  \href{http://dx.doi.org/10.1088/1475-7516/2011/01/030}{{\em JCAP} {\bfseries
  1101} (2011) 030},
\href{http://arxiv.org/abs/1010.3693}{{\ttfamily arXiv:1010.3693 [hep-ph]}}.

\bibitem{Shiu:2011qw}
G.~Shiu and J.~Xu, ``{Effective Field Theory and Decoupling in Multi-Field
  Inflation: An Illustrative Case Study},''
  \href{http://dx.doi.org/10.1103/PhysRevD.84.103509}{{\em Phys.Rev.}
  {\bfseries D84} (2011) 103509},
\href{http://arxiv.org/abs/1108.0981}{{\ttfamily arXiv:1108.0981 [hep-th]}}.

\bibitem{Cespedes:2012hu}
S.~Cespedes, V.~Atal, and G.~A. Palma, ``{On the Importance of Heavy Fields
  during Inflation},''
  \href{http://dx.doi.org/10.1088/1475-7516/2012/05/008}{{\em JCAP} {\bfseries
  1205} (2012) 008},
\href{http://arxiv.org/abs/1201.4848}{{\ttfamily arXiv:1201.4848 [hep-th]}}.

\bibitem{McAllister:2012am}
L.~McAllister, S.~Renaux-Petel, and G.~Xu, ``{A Statistical Approach to
  Multi-Field Inflation: Many-Field Perturbations Beyond Slow Roll},''
  \href{http://dx.doi.org/10.1088/1475-7516/2012/10/046}{{\em JCAP} {\bfseries
  1210} (2012) 046},
\href{http://arxiv.org/abs/1207.0317}{{\ttfamily arXiv:1207.0317
  [astro-ph.CO]}}.

\bibitem{Burgess:2012dz}
C.~Burgess, M.~Horbatsch, and S.~Patil, ``{Inflating in a Trough: Single-Field
  Effective Theory from Multiple-Field Curved Valleys},''
  \href{http://dx.doi.org/10.1007/JHEP01(2013)133}{{\em JHEP} {\bfseries 1301}
  (2013) 133},
\href{http://arxiv.org/abs/1209.5701}{{\ttfamily arXiv:1209.5701 [hep-th]}}.

\bibitem{Chen:2006nt}
X.~Chen, M.-x. Huang, S.~Kachru, and G.~Shiu, ``{Observational Signatures and
  Non-Gaussianities of General Single-Field Inflation},''
  \href{http://dx.doi.org/10.1088/1475-7516/2007/01/002}{{\em JCAP} {\bfseries
  0701} (2007) 002},
\href{http://arxiv.org/abs/hep-th/0605045}{{\ttfamily arXiv:hep-th/0605045
  [hep-th]}}.

\bibitem{Kaplunovsky:1993rd}
V.~S. Kaplunovsky and J.~Louis, ``{Model-Independent Analysis of Soft Terms in
  Effective Supergravity and in String Theory},''
  \href{http://dx.doi.org/10.1016/0370-2693(93)90078-V}{{\em Phys.Lett.}
  {\bfseries B306} (1993) 269--275},
\href{http://arxiv.org/abs/hep-th/9303040}{{\ttfamily arXiv:hep-th/9303040
  [hep-th]}}.

\bibitem{Grisaru}
M.~T. Grisaru, W.~Siegel, and M.~Rocek, ``{Improved Methods for Supergraphs},''
\href{http://dx.doi.org/10.1016/0550-3213(79)90344-4}{{\em Nucl.Phys.}
  {\bfseries B159} (1979) 429}.

\bibitem{Kallosh:2004yh}
R.~Kallosh and A.~D. Linde, ``{Landscape, the Scale of SUSY Breaking, and
  Inflation},'' \href{http://dx.doi.org/10.1088/1126-6708/2004/12/004}{{\em
  JHEP} {\bfseries 0412} (2004) 004},
\href{http://arxiv.org/abs/hep-th/0411011}{{\ttfamily arXiv:hep-th/0411011
  [hep-th]}}.

\bibitem{He:2010uk}
T.~He, S.~Kachru, and A.~Westphal, ``{Gravity Waves and the LHC: Towards
  High-Scale Inflation with Low-Energy SUSY},''
  \href{http://dx.doi.org/10.1007/JHEP06(2010)065}{{\em JHEP} {\bfseries 1006}
  (2010) 065},
\href{http://arxiv.org/abs/1003.4265}{{\ttfamily arXiv:1003.4265 [hep-th]}}.

\bibitem{Conlon:2008cj}
J.~P. Conlon, R.~Kallosh, A.~D. Linde, and F.~Quevedo, ``{Volume Modulus
  Inflation and the Gravitino Mass Problem},''
  \href{http://dx.doi.org/10.1088/1475-7516/2008/09/011}{{\em JCAP} {\bfseries
  0809} (2008) 011},
\href{http://arxiv.org/abs/0806.0809}{{\ttfamily arXiv:0806.0809 [hep-th]}}.

\bibitem{Randall}
L.~Randall and R.~Sundrum, ``{Out of this World Supersymmetry Breaking},''
  \href{http://dx.doi.org/10.1016/S0550-3213(99)00359-4}{{\em Nucl.Phys.}
  {\bfseries B557} (1999) 79--118},
\href{http://arxiv.org/abs/hep-th/9810155}{{\ttfamily arXiv:hep-th/9810155
  [hep-th]}}.

\bibitem{Blumenhagen:2009gk}
R.~Blumenhagen, J.~Conlon, S.~Krippendorf, S.~Moster, and F.~Quevedo, ``{SUSY
  Breaking in Local String/F-Theory Models},''
  \href{http://dx.doi.org/10.1088/1126-6708/2009/09/007}{{\em JHEP} {\bfseries
  0909} (2009) 007},
\href{http://arxiv.org/abs/0906.3297}{{\ttfamily arXiv:0906.3297 [hep-th]}}.

\bibitem{BergMarsh}
M.~Berg, D.~Marsh, L.~McAllister, and E.~Pajer, ``{Sequestering in String
  Compactifications},'' \href{http://dx.doi.org/10.1007/JHEP06(2011)134}{{\em
  JHEP} {\bfseries 1106} (2011) 134},
\href{http://arxiv.org/abs/1012.1858}{{\ttfamily arXiv:1012.1858 [hep-th]}}.

\bibitem{Luty2}
M.~A. Luty and R.~Sundrum, ``{Supersymmetry Breaking and Composite Extra
  Dimensions},'' \href{http://dx.doi.org/10.1103/PhysRevD.65.066004}{{\em
  Phys.Rev.} {\bfseries D65} (2002) 066004},
\href{http://arxiv.org/abs/hep-th/0105137}{{\ttfamily arXiv:hep-th/0105137
  [hep-th]}}.

\bibitem{BergConlon}
M.~Berg, J.~P. Conlon, D.~Marsh, and L.~T. Witkowski, ``{Superpotential
  De-Sequestering in String Models},''
  \href{http://dx.doi.org/10.1007/JHEP02(2013)018}{{\em JHEP} {\bfseries 1302}
  (2013) 018},
\href{http://arxiv.org/abs/1207.1103}{{\ttfamily arXiv:1207.1103 [hep-th]}}.

\bibitem{Anisimov}
A.~Anisimov, M.~Dine, M.~Graesser, and S.~D. Thomas, ``{Brane World SUSY
  Breaking},'' \href{http://dx.doi.org/10.1103/PhysRevD.65.105011}{{\em
  Phys.Rev.} {\bfseries D65} (2002) 105011},
\href{http://arxiv.org/abs/hep-th/0111235}{{\ttfamily arXiv:hep-th/0111235
  [hep-th]}}.

\bibitem{Kachru2}
S.~Kachru, J.~McGreevy, and P.~Svrcek, ``{Bounds on Masses of Bulk Fields in
  String Compactifications},''
  \href{http://dx.doi.org/10.1088/1126-6708/2006/04/023}{{\em JHEP} {\bfseries
  0604} (2006) 023},
\href{http://arxiv.org/abs/hep-th/0601111}{{\ttfamily arXiv:hep-th/0601111
  [hep-th]}}.

\bibitem{Kachru}
S.~Kachru, L.~McAllister, and R.~Sundrum, ``{Sequestering in String Theory},''
  \href{http://dx.doi.org/10.1088/1126-6708/2007/10/013}{{\em JHEP} {\bfseries
  0710} (2007) 013},
\href{http://arxiv.org/abs/hep-th/0703105}{{\ttfamily arXiv:hep-th/0703105
  [hep-th]}}.

\bibitem{Douglas:2006es}
M.~R. Douglas and S.~Kachru, ``{Flux Compactification},''
  \href{http://dx.doi.org/10.1103/RevModPhys.79.733}{{\em Rev.Mod.Phys.}
  {\bfseries 79} (2007) 733--796},
\href{http://arxiv.org/abs/hep-th/0610102}{{\ttfamily arXiv:hep-th/0610102
  [hep-th]}}.

\bibitem{Baumann:2008aq}
D.~Baumann {\em et~al.}, ``{CMBPol Mission Concept Study: Probing Inflation
  with CMB Polarization},'' \href{http://dx.doi.org/10.1063/1.3160885}{{\em AIP
  Conf.Proc.} {\bfseries 1141} (2009) 10--120},
\href{http://arxiv.org/abs/0811.3919}{{\ttfamily arXiv:0811.3919 [astro-ph]}}.

\bibitem{Assassi:2012zq}
V.~Assassi, D.~Baumann, and D.~Green, ``{On Soft Limits of Inflationary
  Correlation Functions},''
  \href{http://dx.doi.org/10.1088/1475-7516/2012/11/047}{{\em JCAP} {\bfseries
  1211} (2012) 047},
\href{http://arxiv.org/abs/1204.4207}{{\ttfamily arXiv:1204.4207 [hep-th]}}.

\bibitem{vanTent:2002az}
B.~van Tent, {\em {Cosmological Inflation with Multiple Fields and the Theory
  of Density Fluctuations}}.
\newblock PhD thesis, Utrecht University,
2002.
\newblock

\end{thebibliography}\endgroup

\end{document}